\documentclass[journal]{vgtc}

\ifpdf
  \pdfoutput=1\relax                   
  \usepackage{graphicx}                
  \DeclareGraphicsExtensions{.pdf,.png,.jpg,.jpeg} 
\else
  \usepackage{graphicx}                
  \DeclareGraphicsExtensions{.eps}     
\fi%

\graphicspath{{figures/}{pictures/}{images/}{./}} 

\usepackage{microtype}                 
\PassOptionsToPackage{warn}{textcomp}  
\usepackage{textcomp}                  
\usepackage{mathptmx}                  
\usepackage{times}                     
\usepackage{cite}                      
\usepackage{tabu}                      
\usepackage{booktabs}                  

\usepackage{color}
\usepackage{subfigure}

\def\_{\rule{.3em}{.15ex}}      

\newtheorem{definition}{Definition}
\newtheorem{property}[definition]{Property}
\newtheorem{proposition}[definition]{Proposition}
\newtheorem{lemma}[definition]{Lemma}
\newtheorem{theorem}[definition]{Theorem}
\newtheorem{corollary}[definition]{Corollary}

\newcommand {\mymarginpar}[1]{\marginpar{#1}}
\renewcommand {\marginpar}[1]{}

\newcommand {\rfig}[1]{Figure \ref{fig:#1}}

\newcommand {\bsec}[2]{\section{#1}
                       \label{sec:#2} }

\newcommand {\rsec}[1]{Section \ref{sec:#1}}

\newcommand {\bsubsec}[2]{\mymarginpar{sec:#2}
                       \subsection{#1}
                       \label{sec:#2} }

\newcommand {\rsubsec}[1]{Section \ref{sec:#1}}

\newcommand {\bsubsubsec}[2]{\mymarginpar{sec:#2}
                       \subsubsection{#1}
                       \label{sec:#2} }

\newcommand {\beq}[1]{
                      \begin{equation}
                      \label{eq:#1} }

\newcommand {\beqno}[1]{\begin{eqnarray}
                      \nonumber}

\newcommand {\eeq}{\end{equation}}
\newcommand {\eeqno}{ && \end{eqnarray}}
\newcommand {\req}[1]{Eq.~(\ref{eq:#1})}

\newcommand {\bear}[1]{
                       \begin{eqnarray}
                       \label{eq:#1} }

\newcommand {\bearno}[1]{
                       \begin{eqnarray}
                       \nonumber}

\newcommand {\eear}{\end{eqnarray}}
\newcommand {\eearno}{\end{eqnarray}}

\newcommand {\btab}[1]{
                       \begin{table}
                       \centering
                       \begin{tabular}{#1}}
\newcommand {\etab}[3] {
                       \end{tabular}
                       \caption[#3]{#2}
                       \label{tab:#1}
                       \end{table}
                       \vspace{.1in}}
\newcommand {\rtab}[1]{Table \ref{tab:#1}}

\newcommand {\btabular}[1]{\begin{center}
                       \begin{tabular}{#1}}
\newcommand {\etabular}{\end{tabular}
                       \end{center}}

\newcommand {\bdefin}[1]{\begin{definition}\label{def:#1}}
\newcommand {\edefin}       {\end{definition}}

\newcommand {\bpro}[1]{\begin{property}
                      \label{pro:#1} }
\newcommand {\epro}   {\end{property}}

\newcommand {\bprop}[1]{\begin{proposition}
                      \label{prop:#1} }
\newcommand {\eprop}       {\end{proposition}}

\newcommand {\blem}[1]{\begin{lemma}
                      \label{lem:#1}}
\newcommand {\elem}   {\end{lemma}}

\newcommand {\bthe}[1]{\begin{theorem}
                      \label{the:#1} }
\newcommand {\ethe}   {\end{theorem}}

\newcommand {\bcor}[1]{\begin{corollary}
                      \label{cor:#1} }
\newcommand {\ecor}   {\end{corollary}}

\newcommand{\hide}[1]{}

\newcommand {\shil}[1]{{\color{red}[From Lei: #1]}}

\vgtcinsertpkg

\title{UrbanFACET: Visually Profiling Cities from Mobile Device Recorded Movement Data of Millions of City Residents}

\author{Lei Shi, Tao Jiang, Ye Zhao, Xiatian Zhang and Yao Lu}
\authorfooter{
\item
 Lei Shi and Tao Jiang are with the State Key Laboratory of Computer Science, Institute of Software, Chinese Academy of Sciences. Email: \{shil,jiangt\}@ios.ac.cn.
\item
 Ye Zhao is with Department of Computer Science, Kent State University. Email: zhao@cs.kent.edu.
\item
 Xiatian Zhang and Yao Lu are with Beijing Tendcloud Tianxia Technology Co., Ltd. Email: \{xiatian.zhang,yao.lu\}@tendcloud.com.
}

\teaser{
\centering
\includegraphics[width=6 in]{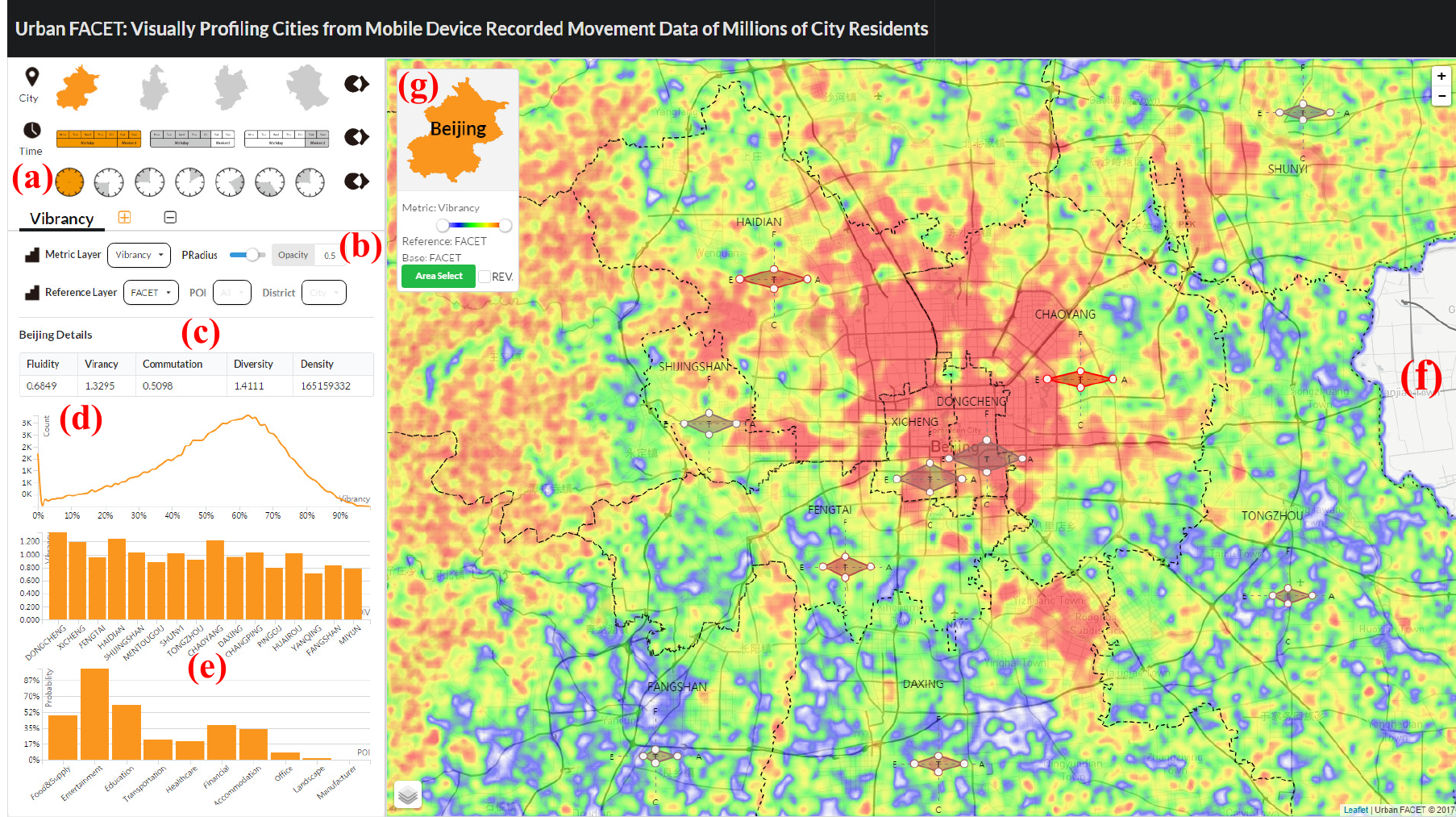}
\vspace{-0.1 in}
\caption{The visualization interface of UrbanFACET: (a) the selection on city and time; (b) the selection of mobility metrics displayed on the map, and the visualization options; (c) detailed information on the selected region (the whole city by default if nothing is selected); (d) probability density function of the displayed mobility metric; (e) average probability distribution on POI/DIV classes to compute the mobility metric; (f) the spatial visualization of mobility metrics (vibrancy in this case); (g) legend and controller of the map.}
\label{fig:UrbanFACETOverview}
}

\abstract{Cities are living systems where urban infrastructures and their functions are defined and evolved due to population behaviors. Profiling the cities and functional regions has been an important topic in urban design and planning. This paper studies a unique big data set which includes daily movement data of tens of millions of city residents, and develop a visual analytics system, namely UrbanFACET, to discover and visualize the dynamical profiles of multiple cities and their residents. This big user movement data set, acquired from mobile users' agnostic check-ins at thousands of phone APPs, is well utilized in an integrative study and visualization together with urban structure (e.g., road network) and POI (Point of Interest) distributions. In particular, we novelly develop a set of information-theory based metrics to characterize the mobility patterns of city areas and groups of residents. These multifaceted metrics including Fluidity, vibrAncy, Commutation, divErsity, and densiTy (FACET) which categorize and manifest hidden urban functions and behaviors. UrbanFACET system further allows users to visually analyze and compare the metrics over different areas and cities in metropolitan scales. The system is evaluated through both case studies on several big and heavily populated cities, and user studies involving real-world users.}

\CCScatlist{ A, B, C
}

\begin{document}


\maketitle


\bsec{Introduction}{Intro} 

%
%
%
%

%


This paper studies the visual analysis of a new kind of human movement data, acquired by synthesizing urban mobile device user's agnostic check-ins from thousands of mobile applications (see details in \rsubsec{Data}). In comparison to previously available human movement data sets from geo-tagged social media or individual mobile apps, our human movement data possesses three unique features. First, it is immensely \textit{huge} because mobile user check-ins are implicitly collected whenever location service is enabled, whereas the check-ins on social media (e.g., Twitter) require users to report by themselves. For example, there are only about 4-5\% tweets are geo-tagged so the achieved trajectories are spatially sparse. Second, the numerous mobile apps cover a wide range of usage scenarios so that our data assembles a \textit{comprehensive} sampling of mobile users' everyday mobility. On the other hand, a single app like traffic map only reports users' check-ins in specific areas (e.g., on the road). Third, the full \textit{context} of user movement is collected as mobile devices (e.g., phones) are probably used exclusively. In comparison, many social media accounts are shared among different users.

A huge amount of data from tens of millions mobile device users in big cities becomes available and demand extensive visual analysis to profile the cities and their citizens. Visualizing such big data of urban human movement opens doors for fine-grained user-level mobility analysis in the metropolitan scale. While techniques such as video surveillance and systematic census could partly profile a urban population map at a high cost, none of them are able to give a clue to ``big'' questions such as: What are the mobility features of the population in city districts? Are there more local residents or casual visitors? How urban areas can be distinguished in population mobility although they may be expected or planned to have the same urban functions? And how the functions of urban areas dynamically vary from day to night, and among workdays and weekend? Answering these questions can visually profile cities and their functional regions as well as the residents, which bring myriad value to (1) urban planners and administrators with enhanced situation and risk awareness for urban planning, management, and public security, and (2) a large spectrum of urban businesses, such as choosing development locations for residential and business real estates; finding optimal places for commercial billboards; and so on.

In this paper, we aims to provide a visual analytics approach of such data to address the following problems: (1) How to extract, measure, and then visualize specific mobility behaviors of cities and residents by correct and meaningful metrics? (2) Can the new metrics that profile city mobility be effectively displayed in together with abundant urban information including map-based context and the classic density map of population? (3) What interactions should be applied to help analyzers quickly discover and understand the mobility metrics? as well as allow them to easily compare among cities and regions? Solving these problems pose multifaceted challenges ranging from data processing, algorithmic analysis and visualization design. 




We propose an integrated visual analysis approach called UrbanFACET which provides effective solutions to these challenges. In particular, the key contributions include: 
\begin{itemize}
 \setlength\itemsep{0em}
 \item We invent a class of information-theoretic metrics to characterize mobility patterns of city regions and city residents. The metrics are based on the fundamental concept of Shannon entropy \cite{shannon2001mathematical}. These entropy-based measures are computed from the distribution of user-level movement data over the spatio-temporal dimension and the categories of urban functions (e.g., shopping, food, business, etc.). Moreover, a decomposition of user entropy leads to a new definition of check-in record entropy that can be aggregated to profile the mobility features of spatial regions. Our data analysis framework is fully optimized for efficiently handling the big movement data. (\rsec{Entropy})
 \item We design and implement the UrbanFACET system which visualizes the entropy-based mobility metrics in together with the record density distribution in a metropolitan scale. Versatile interactions are integrated which allow users to select, filter, and compare the urban mobility patterns in any regions across multiple big cities and in different time periods. The visualization of composite mobility metrics are also studied for specific urban analysis tasks. (\rsec{Vis})
 \item We design efficient data structures and algorithms to process and manage the big urban movement data up to terabytes, so as to well support prompt data access and interactive visualization. Furthermore, we optimize the visualizations specifically for the massive movement data to promote easy understanding and reduce clutters. (\rsec{ScalableComputation}) 
\item The proposed visual analytics system and mobility metrics are evaluated through both case studies on big cities and user studies involving the real-world usage of our system. Usefulness of the system is positively demonstrated by domain users. We also receive constructive comments from urban analytics experts in improving the system. (\rsec{Eva})
\end{itemize}

To our knowledge, this is the first study of its kind to visually profile the mobility of Chinese cities using billions of urban movement data.

\section{Related Work}




In urban studies, exploring patterns and trends of intra-urban human mobility advances the understanding of urban dynamics and reveals socioeconomic driving forces \cite{Chowell03,Sang11,LiuKGXT12}. Location-aware devices are widely applied in urban studies \cite{Pavithra14,Phithakkitnukoon10,Ratti06,Shoval08}. Location data from cell phones is used in urban analysis in Milan \cite{Ratti06} and Rome \cite{Sevtsuk10}, Italy. The increasing availability of GPS data has greatly facilitated the study of street networks \cite{Jiang09,JiangLiu09,WangHB12}. The floating car technique has been used by intelligent transportation systems to obtain its positional information \cite{DaiFR03,TongMC09}. Hence, taxis often serve as floating cars to obtain human mobility data and examine real-time traffic status and individual behaviors \cite{Jiang09,LiuAR10,LiZW11,QiLLPWZ11}.

In the field of data mining, trajectories of human and vehicle motions are used to discover knowledge from large-scale datasets \cite{Zheng:2014:UrbanComputing}. Vehicle trajectory data has been used in traffic monitoring and prediction \cite{Pan:2013:CST}, urban planning \cite{Zheng:2011:UCT}, driving routing \cite{Yuan:2010:TDD,Yuan:2013:TED}, extracting geographical borders \cite{GeoBorder12}, service improvement \cite{Yuan:2013:TRS}, energy consumption analysis \cite{Zhang:2013:SPU}, and dynamic travel time estimation \cite{Pfoser:2008:DTT}. Large-scale mobile phone data with GIS information is used to uncover hidden patterns in urban road usage \cite{WangNature12}, find privacy bounds of human mobility \cite{privacymobility13}, estimate travel time \cite{Thiagarajan:2009:VAE} and infer land use \cite{Toole:2012:ILU}. Public transit trajectories are used in bus arrival time predictions \cite{Zimmerman:2011:FTT}, user's transportation mode inference \cite{Zimmerman:2011:FTT}, and travelers' spending optimization \cite{Lathia:2011:MMD}.

A large number of approaches have been proposed to visually explore movement data (see \cite{andrienko2012visual} for a recent survey). Many of them are focused on the origins and destinations of the trajectories, such as flow maps \cite{phan2005flow}, Flowstrates \cite{boyandin2011flowstrates}, origin-destination (OD) maps \cite{wood2010visualisation}, and visual queries for origin and destination data \cite{Ferreira:2013:VEB}. Other work visualizes trajectories using various visual metaphors and interactions, such as GeoTime \cite{Kapler:2005:GTI}, TripVista \cite{Guo:2011:TTP}, FromDaDy \cite{Hurter:2009:FSA}, vessel movement \cite{Willems:2009:VVM}, route diversity \cite{liu2011diverse}, Kohonen map \cite{schreck2009visual} and more.

Many visual analytics approaches have also been developed to analyze human mobility behaviors. Bicycle hire patterns are discovered by flow maps and OD maps \cite{Wood11}. They were also studied by aggregating individual OD data according to trip directions and distance \cite{MassAbstraction17}. Bluetooth based OD data is also used to find the dynamics of vehicles to characterize urban networks \cite{Laharotte15}. Smart-card data records human behaviors which is used for extracting passenger routes \cite{Hurk15} and analyzing subway routes and reachable regions \cite{zeng2014visualizing}. Social media data with geo-tags \cite{GeoTweet13,GeoPhoto10} can also find places and events related to urban human movement. However, none of the existing work has been utilizing the massive mobile device recorded data as in this paper, which includes tens of millions of real users in several big cities. 

The geographical context of cellular towers and an alternative modularity function is used to interpret  the  patterns  in  the  phone  call interactions and the  mobile phone users’ movement \cite{Gao13}. The cellular data network records are also used to compute a finer granularity of location and movement \cite{Zhang14}. MViewer system \cite{Pu2014} is designed to visualize and analyze the population mobility patterns from millions of phone call records. It includes visual analysis of user groups in a base station, the mobility patterns on different user groups in certain base stations, and handoff phone call records.
Cell phone location records are used for studying urban human flow across a city with flow volumes, links, and user communities \cite{MobileViewer16}. These approaches do not provide entropy-based metrics for human mobility analysis as we do here.

Entropy metric has been used by Song et al. \cite{Science10} to study the mobility patterns of anonymized mobile phone users. They used 50,000 individuals anonymous mobile phone users in a 3-month-long record. Their major goal was to find to what degree human behavior is predictable, which was completely different from ours. By measuring the entropy of each individual’s trajectory, they found a 93\% potential predictability in user mobility. This work showed that population mobility patterns might be discovered through analyzing mobile users data, which partially justifies our approach in this work. Recently, Kang et al. \cite{Wuhan17} used mobile records to uncover frequently visited locations of a city's massive mobile users where an approximate entropy is used. This entropy definition was different from our metrics and they did not provide visual systems of several big cities.


\bsec{Urban Data Source and User Task}{DataTask}

\begin{figure*}[t]
\centering
\includegraphics[width=7 in]{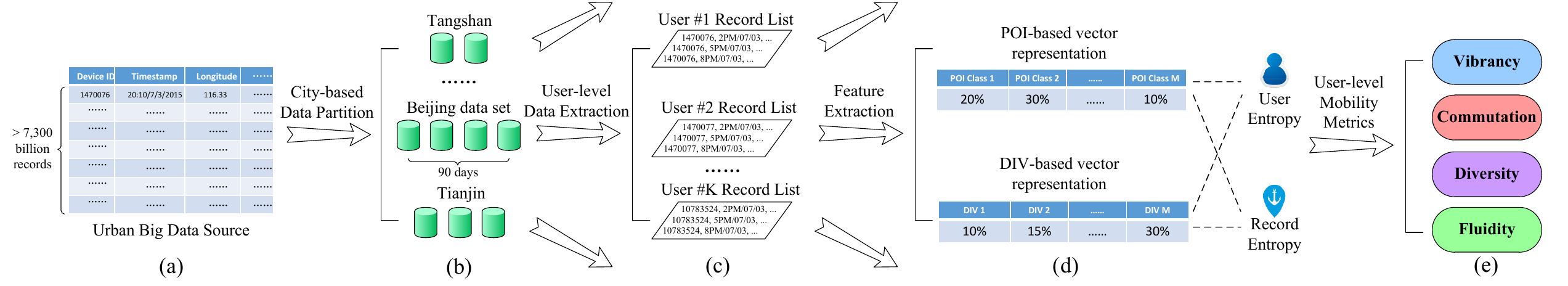}
\vspace{-0.1 in}
\caption{The user-level mobility analysis pipeline of UrbanFACET.} 
\vspace{-0.1 in}
\label{fig:AnalysisPipeline}
\end{figure*}

\bsubsec{Data Description}{Data}

%

%
%
%

\begin{table}[t]
\centering
\caption{The metadata of location records in our data source.}
\label{tab:DataFormat}
\begin{tabular}{|l|l|l|}
\hline
Field                            & Description   & Sample   \\ \hline \hline
Time      & Timestamp of the record & 20:10/07/03/2015   \\ \hline
Lon.      & Longitude of location & 116.3336266             \\ \hline
Lat.      & Latitude of location & 39.890955             \\ \hline
Mid       & Unique ID of the device & 1470076020481    \\ \hline
Src       & Source of the location record & GPS \\ \hline
%
\end{tabular}
\vspace{-0.2 in}
\end{table}

\begin{table}[t]
\centering
\caption{Statistics on four data sets used in this research.}
\label{tab:DataCollection}
\begin{tabular}{|l|l|l|l|l|}
\hline
City        & \#Device      & \#Record       & Size      & Time                      \\ \hline \hline
Beijing     & 31849742    & 8407648917     & 738.1G    & 90 days                \\ \hline
Tianjin     & 8011128     & 2858575880     & 206.8G    & 90 days                      \\ \hline
Tangshan    & 2786668     & 920364499      & 64.8G     & 90 days                      \\ \hline
Zhangjiakou & 1392236     & 317252149      & 23.1G     & 90 days                    \\ \hline
\end{tabular}
\vspace{-0.05 in}
\end{table}

The big urban movement data is provided by a mobile analytics company which keeps the real-time tracking of billions of smart devices in China, including mobile phones, tablets, wearable devices, etc. This data is collected by registering third-party APIs inside the mobile apps of each device. When a registered mobile app is activated (not necessarily being used), the API will report location records to the company server continuously. The metadata of location records is shown in \rtab{DataFormat}. There are four fields in each record: the time of recording, location information (longitude and latitude), unique ID of the smart device, and the localization method (i.e. source, including GPS, Wi-Fi, base station and Internet IP). Here is a sample record:``20:10/07/03/2015, 116.3336266,39.890955,1470076020481,GPS''.


Compared with previously published urban movement data sets \cite{li2008mining,miranda2017urban}, our data source has the following compelling features that empower the work here. First, the volume of our movement data is \textit{huge} because the third-party API has been installed with more than 120,000 kinds of mobile apps, covering a majority of China's 594 million active mobile devices \cite{NumOfActiveDevice2015}. The incoming records are averaged to 4 billions per day in total, and more than 100 millions per day for each of the top ten Chinese cities. Second, the composition of these location records is \textit{comprehensive} in that the 120,000 mobile apps involve a wide spectrum of application domains, such as entertainment, education, information, etc. This is unlike the data collected from a single app of traffic map which mostly records user locations on the road or transportation lines. Third, the \textit{user context} can be extracted according to the unique device ID in an unambiguous way. This opens the door for fine-grained user-level urban movement data analysis.


In this work, we conduct analysis on four representative data sets extracted from the company's data repository. As listed in \rtab{DataCollection}, each data set corresponds to a 90-day collection of one Chinese city, including Beijing (capital of China), Tianjin (one of five national central cities in China), Tangshan and Zhangjiakou (two major cities in Hebei province surrounding Beijing and Tianjin). These four cities form the so-called national capital region of China. To obtain more accurate analysis results, we only keep the records collected by GPS and Wi-Fi, from which sources the spatial precision of records is kept below 100 meters. The final data sets are relatively smaller than the raw data, with the largest city (Beijing) having 93.4 million records per day. User privacy is preserved by anonymizing the device ID and discretizing time. For each timestamp, only the 10-minute interval it belongs to is kept.

\bsubsec{Task Characterization}{Task}


With the big urban movement data enhanced by precise user context, it is possible for us to carry out fine-grained, user-level visual analysis to profile cities, their functional regions, and group of residents in the metropolitan scale. The city profiling result can help government, enterprise and individuals in various aspects, including city planning, site configuration, situation and risk awareness. In details, we target three user-level mobility analysis tasks in this work:
\begin{itemize}
 \setlength\itemsep{0em}
\item \textbf{Overview} of multidimensional user mobility metrics on the geospatial city map under certain time frame. City administrators can gain situation and risk awareness by examining the spatiotemporal distribution of their residents' mobility.
\item \textbf{Correlation analysis} of the distribution of user mobility metrics with spatial Point of Interests (POI), regional demographics (e.g., GDP) and population density. Enterprise and individuals can make their business decisions (e.g., site configuration, real estate investment) based on the analysis and reasoning of local user mobility metrics.
\item \textbf{Comparison analysis} among cities and time periods (e.g., morning and night) to reveal the city-level mobility disparity and potential temporal patterns. Composite mobility metrics can also be studied to compare between city functional regions. Government can optimize their city planning, including land and transportation management, based on the comparison result.  
\end{itemize}
\bsec{User-Level Mobility Analysis}{Entropy}



\bsubsec{User-based Feature Extraction}{UserExtraction}


The big urban movement data is analyzed by a pipeline given in \rfig{AnalysisPipeline}. The pipeline takes the raw data described in \rsubsec{Data} as input, which sums up to more than 7,300 billion records (\rfig{AnalysisPipeline}(a)). In the first step, the big data is partitioned according to the record location into cities by their administrative boundary (\rfig{AnalysisPipeline}(b)). This city-based partition seems straightforward, but the actual implementation is nontrivial, which involves the challenging task of big data query. In this work, we focus on visual analytics part and leverage the data infrastructure provided by the mobile analytics company for the query operation.

We obtain a same 90-day data set (2015) on four Chinese big cities respectively, as shown in \rtab{DataCollection}. Take Beijing as an example, in the next step, all the records are aggregated by their device ID for user-level data extraction (\rfig{AnalysisPipeline}(c)). The underlying assumption is that each device is linked to a unique mobile user, so that the record list of each device can reveal the mobility pattern of the corresponding user. To satisfy this assumption, a follow-up data cleansing step is employed. All the devices having smaller than one records per month or larger than 2,500 records per month are removed. The over low record number suggests that the device may not be the primary one for its owner or the owner is not a local citizen. On the other hand, the over high record number mostly leads to an abnormal usage/type of the device, such as robots, fixed sensors. 
The effective window of [1, 2500] is selected based on the domain knowledge of our collaborator in the mobile analytics company. Throughout this work, ``user'' will be used interchangeably with ``device''.

On each user, we obtain a list of records with timestamp and longitude/latitude location. To analyze user-level mobility, the first question is how to represent this movement data for each user. A straightforward method is to build a location vector and a time vector which use each exact location and time as columns and the number of user occurrence as the cell value. In this vector-based representation, the user by location and user by time matrix can be constructed. However, in reality most such matrices will be extremely sparse, making it hard to perform classical data analysis such as clustering and classification.

\begin{figure}[t]
\centering
\includegraphics[width=3.5 in]{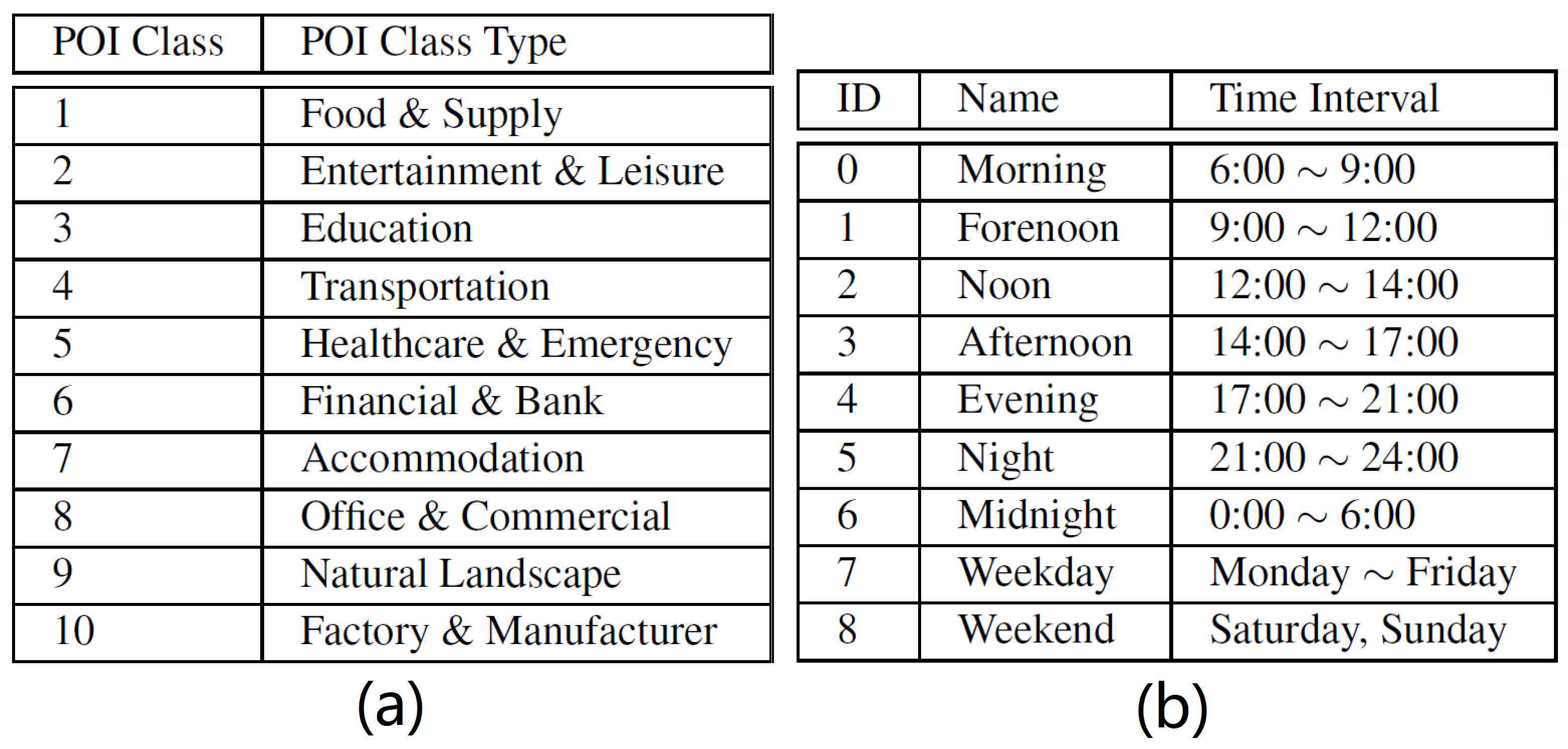}
\vspace{-0.3 in}
\caption{The list of (a) POI classes and (b) time interval divisions used in the feature extraction.}
\vspace{-0.1 in}
\label{fig:DataTable}
\end{figure}

\begin{figure}[t]
\centering
\includegraphics[width=2.5 in]{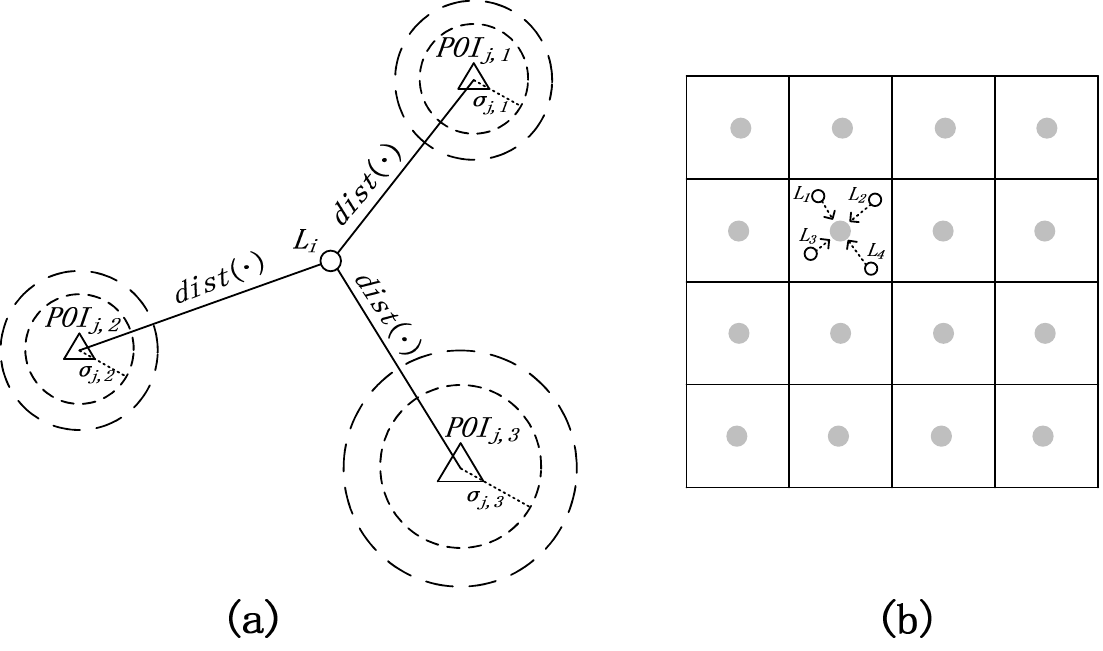}
\vspace{-0.1 in}
\caption{Computing POI class probability for each record: (a) Sum of multiple Gaussian probability density function; (b) the scalable computation method by using grid-based map created from the square lattice.}
\vspace{-0.1 in}
\label{fig:POIClassComputation}
\end{figure}

In this work, we exploit the spatial and temporal context of each urban movement record by feature extraction. Using the extracted features, a dense representation of each user can be computed. On the spatial aspect, we consider two kinds of features. The first is the Point of Interests (POI) associated with the location record. \rfig{DataTable}(a) lists 10 classes of POIs used in this work, adapted from OpenStreetMap POI types \cite{OpenStreetMap}. For each user with a list of $N$ records $\{R_i\}_{i=1,\cdots,N}$. The location of the $i$th record $R_i$ is denoted as $L_i$. We compute the probability of $R_i$ belonging to the $j$th POI class by
\beq{POIClassComputation}
q_{ij}=\frac{\sum_{k}{\varphi_{0,\sigma_{j,k}^2}(dist(L_i,POI_{j,k}))}}{\sum_{j}\sum_{k}{\varphi_{0,\sigma_{j,k}^2}(dist(L_i,POI_{j,k}))}}
\eeq
where $\varphi_{0,\sigma_{j,k}^2}$ indicates the probability density function of a Gaussian distribution with zero mean and a variance of $\sigma_{j,k}$. $dist(L_i,POI_{j,k})$ indicates the spatial distance between the record's location $L_i$ and the $k$th POI of class $j$ ($POI_{j,k}$). An illustration is given in \rfig{POIClassComputation}(a). The probability is summed from the Gaussian-distributed influence of each $j$th-class POI, before normalized across all POI classes. Note that only POIs close to $L_i$ are considered, by default the valid range is set to 500 meters. If there is no $j$th-class POI close to $L_i$, we set $q_{ij}=0$. $\sigma_{j,k}$, the influence variance of the $k$th POI of class $j$ ($POI_{j,k}$), is set to 1.5 times of the radius of $POI_{j,k}$ if the POI is area-based, or 100 meters if the POI is point-based. All these parameters come from the practical experience of data scientists in the mobile analytics company.

The second kind of spatial feature is the city's administrative division (DIV) of each location record. Because DIVs are mostly non-overlapping, the feature extraction is straightforward, i.e., $q_{ij} = 1$ or $0$ for all records, depending on whether or not the location $L_i$ is within the $j$th DIV. The temporal feature takes a similar form by dividing time into non-overlapping intervals. We have two kinds of time divisions, as shown in \rfig{DataTable}(b), ID 0$\sim$6 separate one day into 7 intervals, and ID 7$\sim$8 separate one week into weekdays and weekend.

After the feature extraction at each record, we aggregate these features into dense representations of each user (\rfig{AnalysisPipeline}(d)). Take the POI-based spatial feature as an example, each user can be represented by a vector $\vec{x} = <p_1, \cdots, p_M>$ where
\beq{POIVectorComputation}
p_j=\frac{\sum_{i}{q_{ij}}}{\sum_{j}\sum_{i}{q_{ij}}}
\eeq
$p_j$ denotes the probability of this user accessing $j$th-class POIs, after normalization across all the $M$ POI classes to satisfy $\sum_{j}{p_j}=1$. Apart from $\vec{x}$, each user will also have a DIV-based spatial feature vector representation and a time-based representation.

\bsubsec{Information-Theoretic User Mobility Metrics}{UserMetric}


\begin{figure}[t]
\centering
\includegraphics[width=2 in]{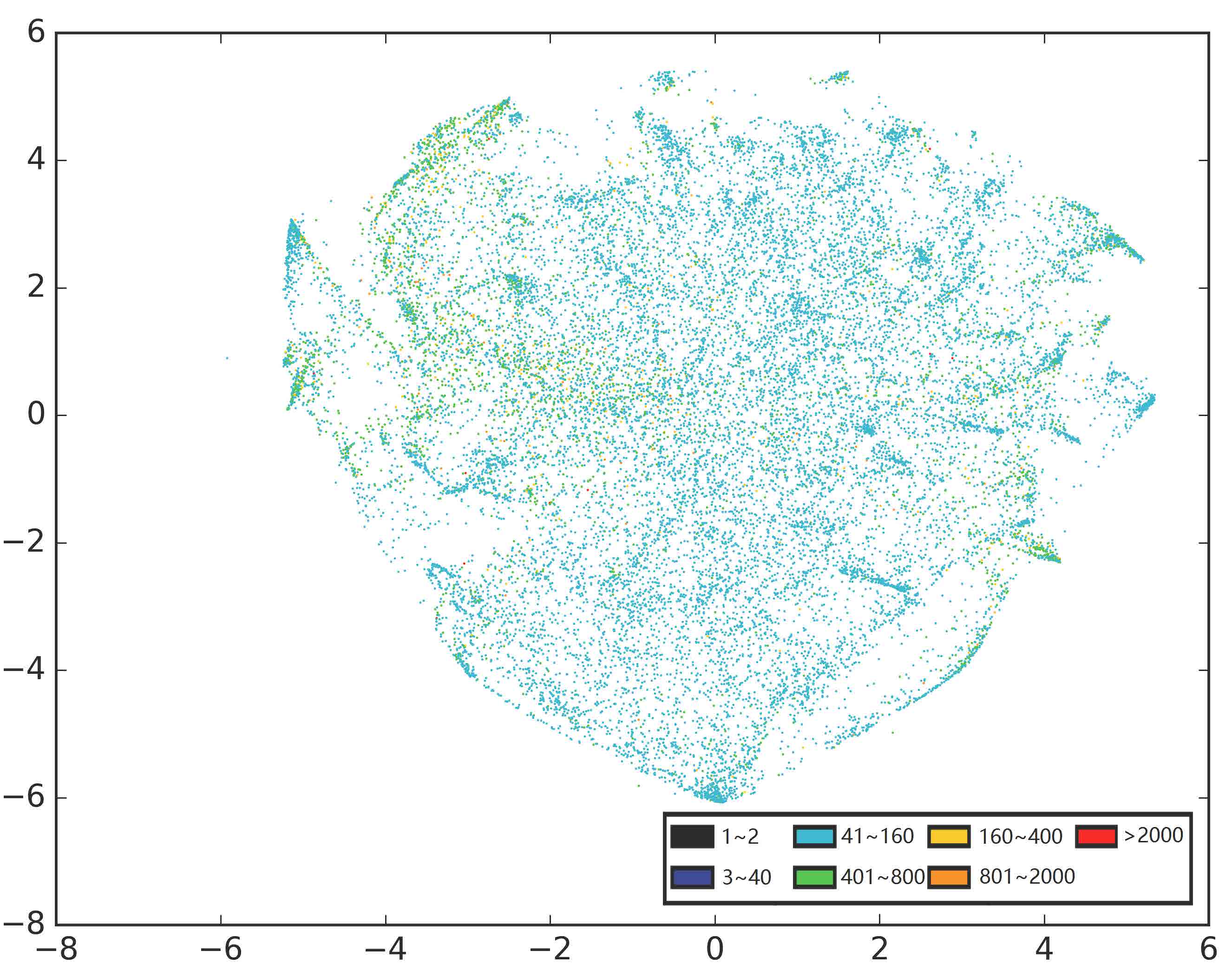}
\vspace{-0.1 in}
\caption{The t-SNE projection result of the user's POI class vectors in a $0.1\%$ random sample of the Beijing data set. The point color in the projection indicates the number of records for each user.}
\vspace{-0.1 in}
\label{fig:UserVectorProjection}
\end{figure}

After the computation of dense vector representation on urban users, we follow-up to study the mapping from these representations to user mobility metrics, so that a map-based visualization can be created for further user mobility analysis. The natural choice is to employ the data clustering to separate different user groups, or use supervised classification based on user category labels. However, after extensive trials, we discover two major difficulties in applying these routine methods. First, in real-life big urban movement data, the user's representation spreads in a sophisticated manner within the vector space such that it is hard to determine both the number of clusters and their clear clustering boundaries. \rfig{UserVectorProjection} depicts the projection result of a random sample of $0.1\%$ users in the Beijing data set, using the POI-based vector representation and t-SNE dimensionality reduction \cite{TSNE}. A larger sample will lead to even worse result. 
The second difficulty lies in defining precise user groups from the clustering result of vector representations. This leads to the incapability of interpreting user clustering result or explicitly labeling users for classification.


Due to the difficulty to discover discrete urban user classes, we consider the idea of defining continuous user mobility metrics over the user-level urban movement data. While existing user-level metrics such as the number of records and total distance of movements are available, the exact value of these metrics depends heavily on the data sampling rate and can be biased for mobility analysis. For example, a frequently recorded user may not indicates s/he is travelling a lot. To get rid of the effect of sampling, we propose to compute user-level metrics based on their probability distributions on spatial and temporal classes, i.e., the vector representation computed in \req{POIVectorComputation}. These distributional metrics will be plausible assuming the sampling is uniform at a reasonable rate. This assumption is satisfied by our data cleansing operation and the data collection method covering a large number of diversified mobile apps ($>120,000$).

In details, we propose to borrow the notion of Shannon entropy \cite{shannon2001mathematical} (or entropy in short) in defining user-level mobility metrics. First, essentially the entropy definition refers to disorder and uncertainty, which is ideal for measuring user's mobility. Second, by information theory, entropy also quantifies the average information acquired in seeing an instance of the random variable, i.e., the spatial/temporal class of a user's record. Mapping user's entropies onto certain spatial region can indicate the information content of this region, thus used for city profiling. Mathematically, take the POI-based vector representation $\vec{x} = <p_1, \cdots, p_M>$ as an example, we define \textbf{user entropy} by
\beq{UserEntropy}
H_p=-\sum_{j=1}^{M}p_{j} \cdot \log p_{j}
\eeq
By default, the natural logarithm is used for the entropy.


To visualize the user entropy on the geospatial map, a key choice is to determine the mobility of each record from their user entropy values. We design two strategies here. The baseline strategy applies the user entropy directly to each record belonging to this user. The resulting map reveals the sum of user mobility appeared in each location, but the variation within a single user's records is discarded. In fact, users can bring more uncertainty (disorder) to their less-visited POI classes. To resolve this deficiency, we compute a decomposition of the sum of each user's entropy, and then try to optimally redistribute this entropy sum according to the record's uncertainty.
\bear{RecordEntropyDerivation}
N \cdot H_p = & -\sum_{j=1}^{M}N \cdot p_{j} \cdot \log p_{j} \nonumber \\
= & -\frac{N}{\sum_{i=1}^{N}\sum_{j=1}^{M}{q_{ij}}} \cdot \sum_{j=1}^{M}\sum_{i=1}^{N}{q_{ij}} \cdot \log p_{j} \nonumber \\
\approx & -\sum_{j=1}^{M}\sum_{i=1}^{N}{q_{ij}} \cdot \log p_{j} \nonumber \\
= & \sum_{i=1}^{N}(-\sum_{j=1}^{M}{q_{ij}} \cdot \log p_{j}) \nonumber
\eear
Based on this decomposition, we define the \textbf{record entropy} by
\beq{RecordEntropy}
H_r=-\sum_{j=1}^{M}q_{ij}\cdot \log p_{j}
\eeq
Compared with user entropy, the metric of record entropy emphasizes more on the property of spatial regions and locations. Take the POI-class based record entropy as an example, a region will have a high average record entropy if most users in this region are casual visitors.


The entropy-based mobility metric definition should be combined with the user's feature vector representation in the actual computation. Below we focus on the spatial feature representations, i.e., the POI-class based and the DIV-class based. The temporal feature representation is not discussed more because they are heavily influenced by the temporal usage of mobile devices, which can be less relevant to the user's mobility pattern. By pairing the feature representation methods with the entropy definition, in this work we invent four user-level mobility metrics (\rfig{AnalysisPipeline}(e)).

\begin{itemize}
    \item \textit{Vibrancy (user entropy over POI class distributions)}: this metric computes the degree of a user switching among different classes of POIs. The vibrancy metric infers the current (or at least potential) richness of the user's personal life. 
    \item \textit{Commutation (user entropy over DIV class distributions)}: this metric computes the degree of a user switching among different administrative divisions. The commutation metric infers the amount of commuting in user's everyday life.
    \item \textit{Diversity (record entropy over POI class distributions)}: this metric computes the scarcity of the record's POI class in its user's spatial region. An area with many records of high diversity means most users there are occasional visitors to this POI class and are potentially different from each other.
    \item \textit{Fluidity (record entropy over DIV class distributions)}: this metric computes the scarcity of the record's administrative division in its user's spatial appearance. A region full of high fluidity records suggests that most users there are occasional visitors to this administrative division. 
\end{itemize}

In spatial visualization, we compute the average mobility metric of all records in a region to profile its mobility. The sum of metrics is not used because it will interfere heavily with the density distribution across regions.

\bsubsec{Scalability Issues}{ScalableComputation}


In computing information-theoretic user mobility metrics, the main challenge is the huge urban data volume. The initial step sorts billions of record according to the device ID, so that mobility metrics can be computed on a per-user basis. We design a file-based indexing system: a hash function is used to map the device ID into one of $B$ smaller record files ($B=10,000$ in our implementation). Each file stores all the records mapped to it, in the order of their device ID. This design enables us to process multiple record files in parallel without I/O conflicts.


In the actual mobility metric computation, the most costly operation is to calculate the probability of each record belonging to each POI class, e.g., $q_{ij}$ for the $i$th record over the $j$th POI class. In theory, we need to fit the distance between each record location and all POIs using their respective Gaussian distribution, which involves a complexity of $O(R \cdot P)$. $R$ is the number of total records and $P$ is the number of all POIs. Though the computation can be optimized by only considering nearby POIs within a radius threshold (500 meters in this work), it is still incapable of computing our Terabyte-level urban big data.


We propose a grid-based solution to the scalability problem. As shown in \rfig{POIClassComputation}(b), we put a point set of square lattice on the map, with an intra-point interval of 200 meters. The Voronoi diagram over this lattice gives a regular tessellation of squares, i.e., the grid-based partition of space with a 100-meter grid radius. According the property of Voronoi diagram, the center of each grid is the closest central point to all the other points in the same grid. Based on this property, we approximate the probability of each location affiliated with POI classes by the probability of the center in the same grid. In this way, we only need to calculate the probability information for all the grid centers. For the city of Beijing, we have about 0.8 million grids, four magnitude smaller than the number of records in our data set. On each record, the processing is reduced to two multiplications on the longitude and latitude to determine which grid it belongs to.


The grid-based solution also makes it easy to visualize user mobility metrics on the map. Instead of depicting each record's mobility separately, we average the user's mobility metric on the same grid and display this average metric onto the map by filling the grid with a uniform color. This allows to separate different record classes in the visualization (e.g., by temporal and contextual information of records). We will give more details in describing the space/time filtering interaction in \rsubsec{MobilityVis}.

\bsec{Visualization Design}{Vis}


%
%
%
%



\rfig{UrbanFACETOverview} presents the UrbanFACET visualization interface. It is composed of three panels. In the top-left visualization configuration panel (\rfig{UrbanFACETOverview}(a)(b)), several options are provided to let user determine what shall be presented in the mobility map on the right (\rfig{UrbanFACETOverview}(f)). Upon a region selection on the mobility map, the details of this region and the associated mobile user profiles will be displayed in the statistics panel (\rfig{UrbanFACETOverview}(c)(d)(e)).

The overall design principle serves the user analysis process to fulfill the three classes of tasks described in \rsubsec{Task}. First, users pick a city and time to investigate (\rfig{UrbanFACETOverview}(a)) and choose to look at one particular mobility metric (\rfig{UrbanFACETOverview}(b)). Then the \textit{overview} distribution of this metric is displayed on the map panel (\rfig{UrbanFACETOverview}(f)). S/he can overlay or juxtapose location references, important regions or another metric to conduct visual \textit{correlation analysis} for reasoning. Details can be obtained in the statistics panel (\rfig{UrbanFACETOverview}(c)(d)(e)) to interpret the visualization result. Finally, a \textit{comparison analysis} can be conducted over city, day of week, or time of day through advanced visualization configurations (\rfig{UrbanFACETOverview}(a)).

In the following, we describe the detailed design on map visualization, interaction and comparison analysis respectively. Several techniques are introduced to solve specific challenges on scalability, usability and visual complexity.



\bsubsec{Mapping User's Mobility Metrics}{MobilityVis}





\begin{figure*}[t]
\centering
\subfigure[]{\includegraphics[height=1.4 in]{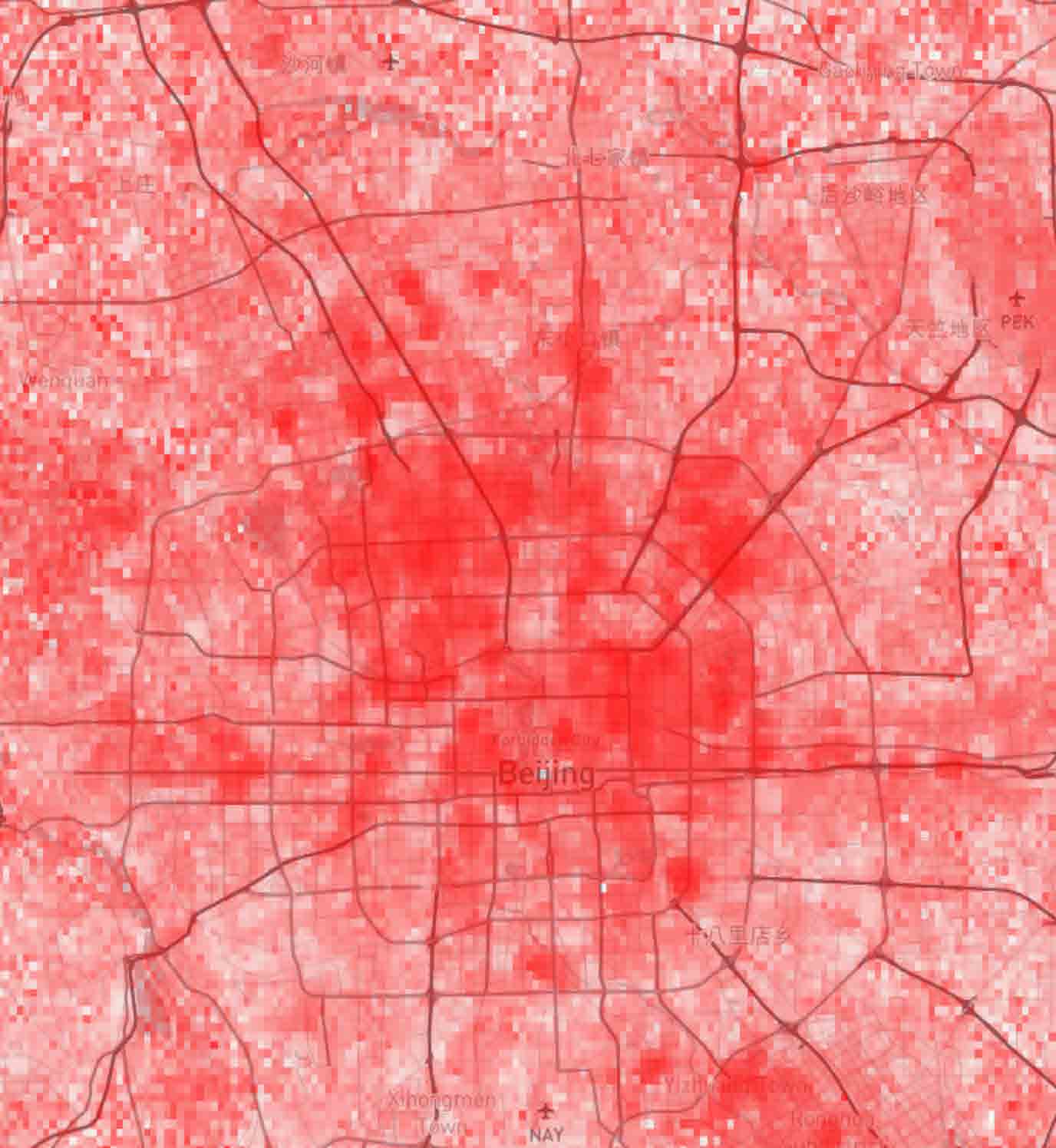}}
\subfigure[]{\includegraphics[height=1.4 in]{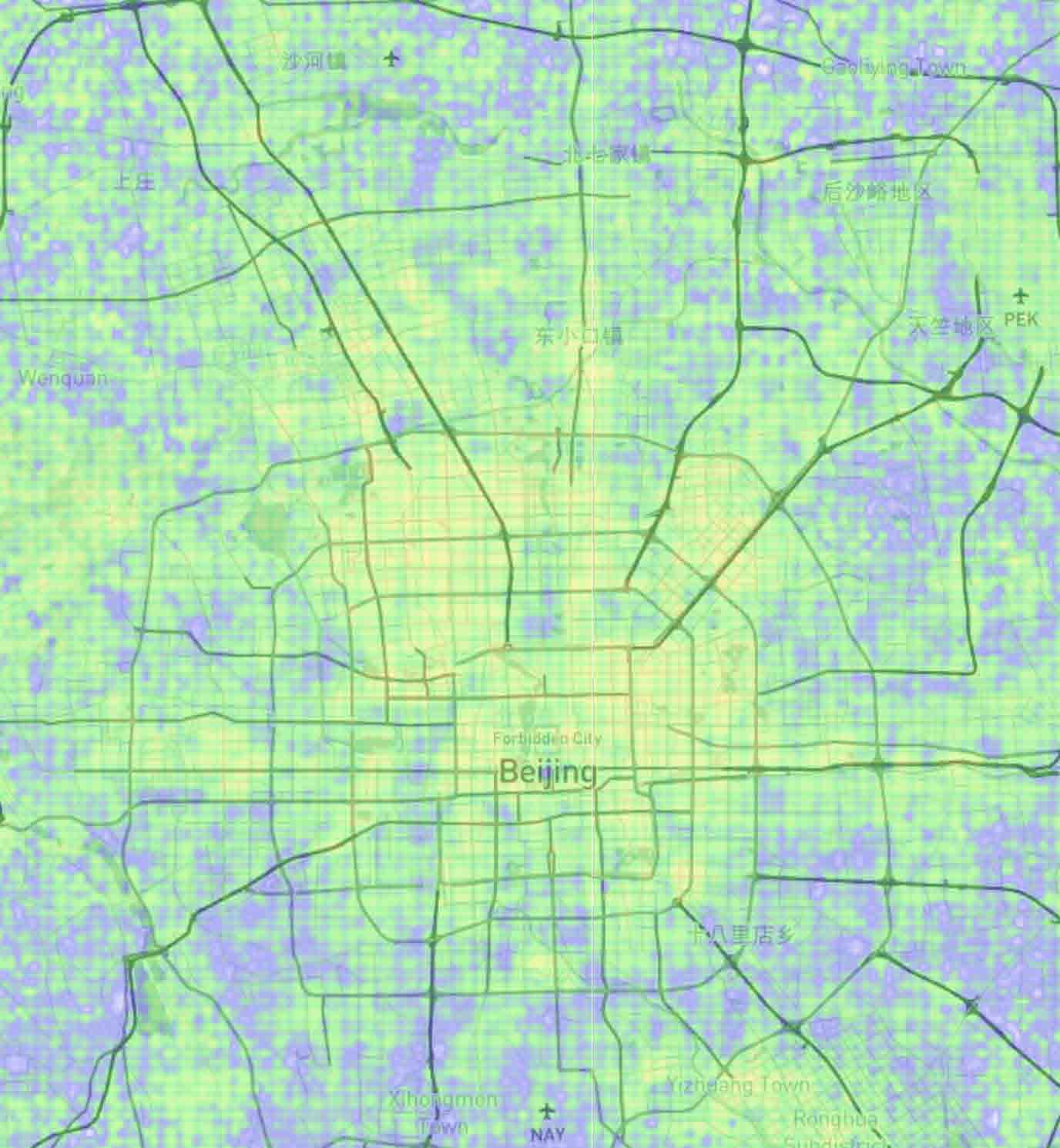}}
\subfigure[]{\includegraphics[height=1.4 in]{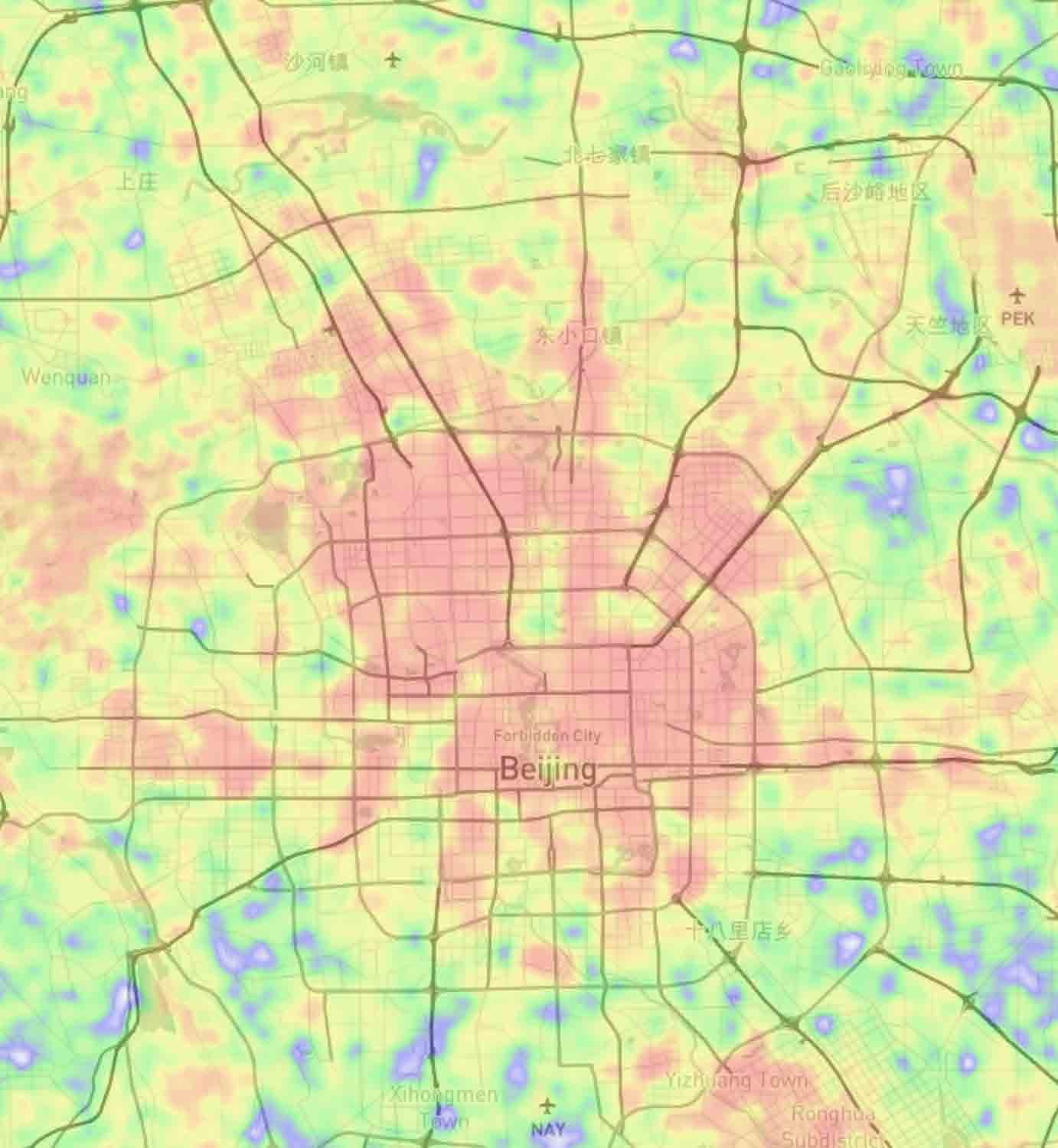}}
\subfigure[]{\includegraphics[height=1.4 in]{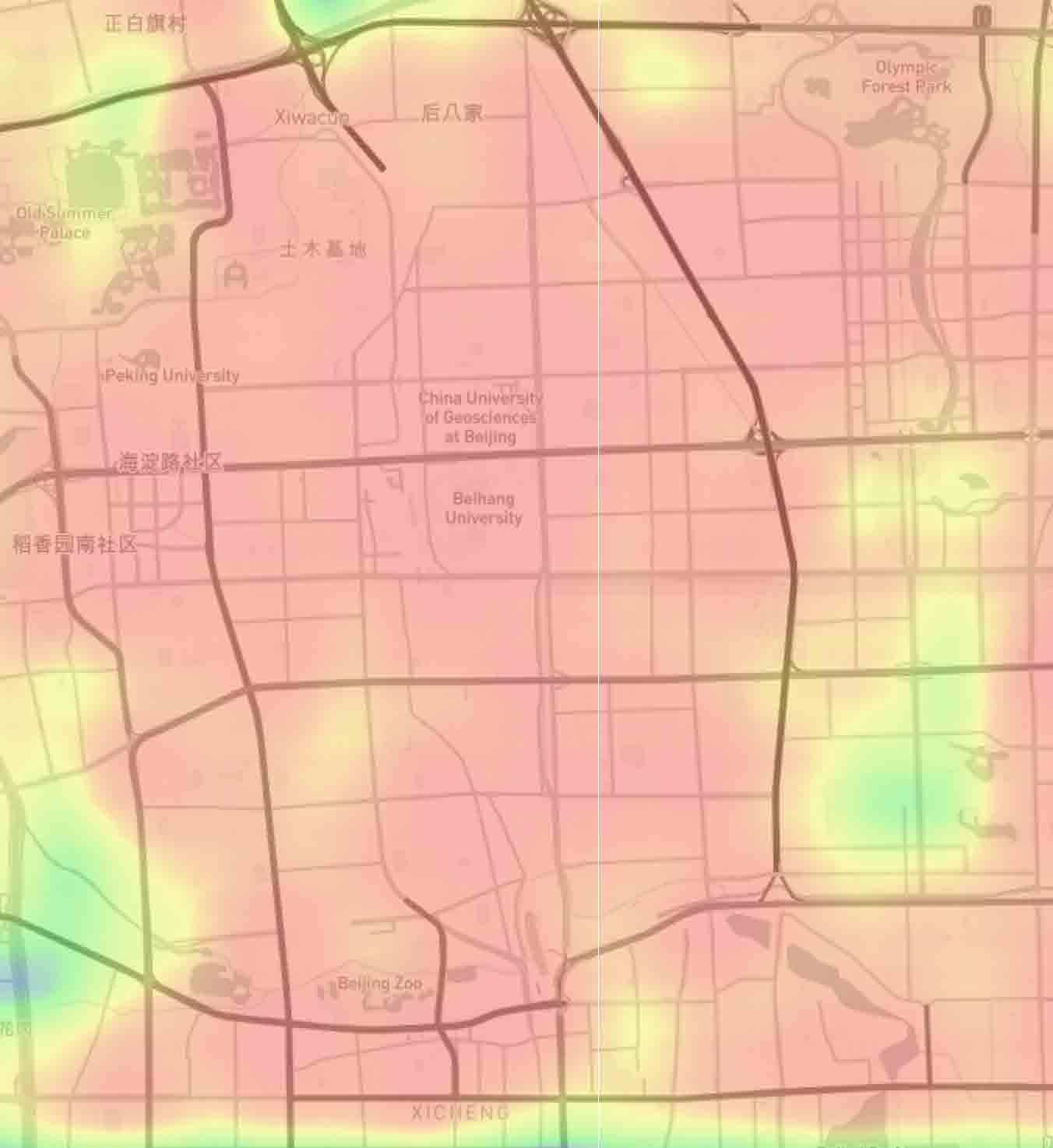}}
\subfigure[]{\includegraphics[height=1.4 in]{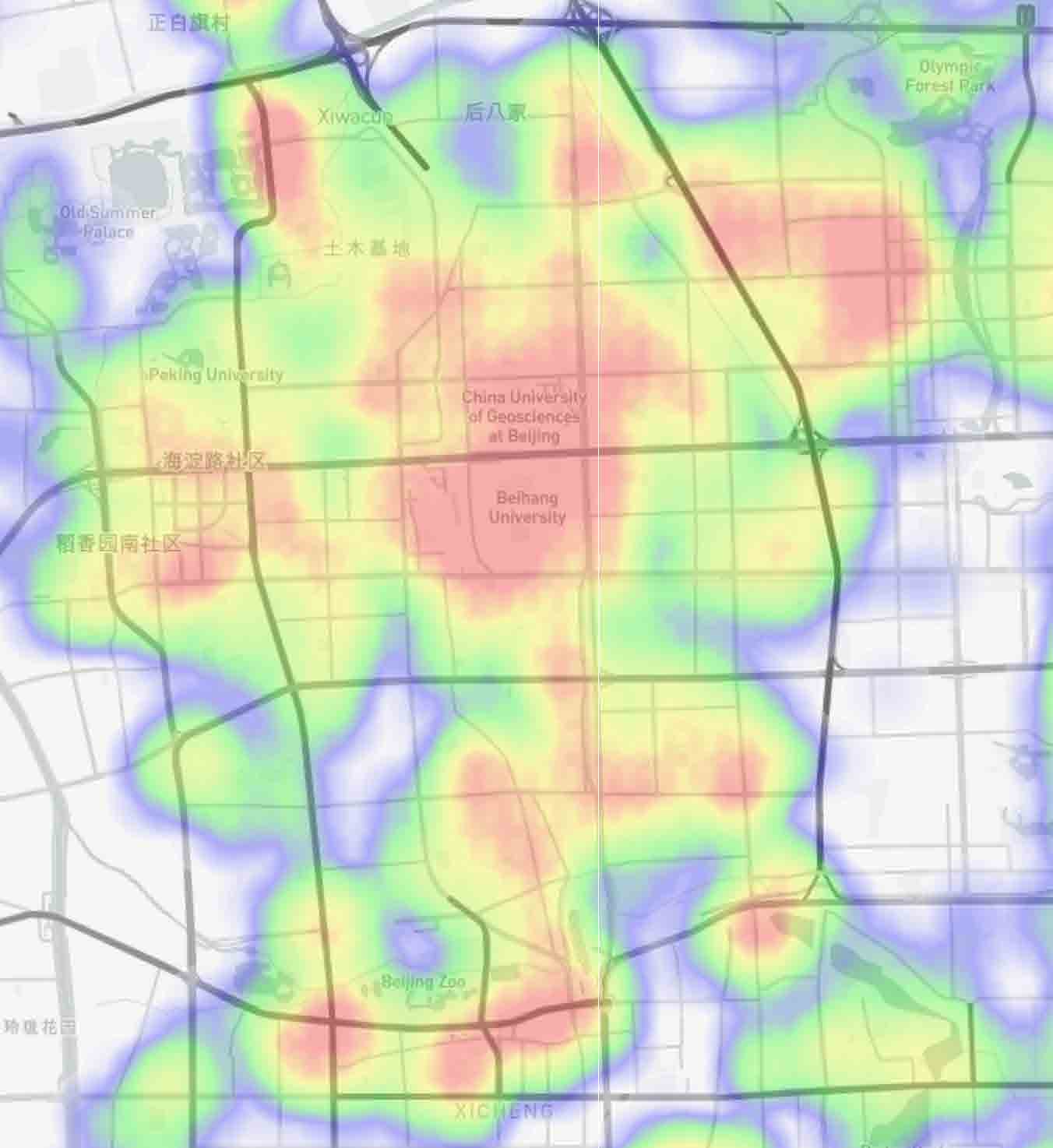}}
\vspace{-0.2 in}
\caption{Alternative mobility metric visualization: (a) grid-based; (b) minimal contour map; (c) contour map with optimal diffusion radius and double color map filters; (d) zoom-in view with fixed diffusion radius; (e) zoom-in view with adaptive diffusion radius.}
\vspace{-0.2 in}
\label{fig:AlternativeDesign}
\end{figure*}

The default view of the map visualization (\rfig{UrbanFACETOverview}(f)) has two layers: the base layer and the metric layer. The base layer in the background gives the geospatial information about the city, serving as location references. We provide multiple choices for the base layer, including road network, terrain, and satellite imaging, which can be used for different correlation analysis scenarios. The metric layer is overlaid on top of the base layer, in adjustable transparency (\rfig{UrbanFACETOverview}(b)), to serve the key role of displaying the mobility metric distribution within a city. In total, four information theoretic mobility metrics (vibrancy, commutation, diversity, fluidity), the average density metric, and three region-based demographics (GDP, population, house price) can be selected in the metric layer visualization. To represent the metric value of each record, we apply a linear mapping from the value to the color hue, which forms a blue to red rainbow palette, as shown in the legend panel of \rfig{UrbanFACETOverview}(g).

The biggest challenge in metric layer visualization lies in scalability. There can be billions of records to map in a single city (e.g., Beijing). Drawing all these records at their accurate locations simultaneously will be impossible, due to the large amount of induced computation/rendering cost and the required visual bandwidth. As introduced in \rsubsec{ScalableComputation}, we have proposed a grid-based solution to the scalability problem. In this solution, the city map is partitioned regularly into small grids. In each grid, the average mobility metric of all records within the grid is used for the representation. This reduces the visual complexity from the number of records (8.41B for Beijing) down to the number of grids (227.9K for Beijing).

A straightforward visual design for the grid-based solution is to fill each grid with the color mapped from the average mobility metric, as shown in \rfig{AlternativeDesign}(a). This design displays the spatial distribution of mobility metrics from raw data, but has two obvious drawbacks. First, the discrete grid partition features incontinuity and can interrupt user's perception on trends, hotspots and valleys. Second, the fixed boundary on grids can distort the actual spatial distribution of mobility metrics. These drawbacks can be alleviated by using finer grid partitions, which however elevate the visual and computational complexity. In this work, we adopt the design of point diffusion based contour maps. The example in a minimal diffusion setting is given in \rfig{AlternativeDesign}(b). On each grid, we place a seed point at the grid center, which diffuses the average mobility metric value from the center in radial directions. The gradient of spatial diffusion is set to constant and the maximal diffusion range is controlled by an diffusion radius setting. The influence of multiple grids is summed up in overlapping areas, so that the discreteness by the grid partition gets smoothed. The resulting visualization forms a contour map if an appropriate diffusion radius is used (\rfig{AlternativeDesign}(c)).

There are several options to choose the diffusion radius. The default method use the fixed radius, e.g., three times of the grid radius, as shown in \rfig{AlternativeDesign}(c). The problem with the fixed diffusion radius is, when users zoom-in the map, the resolution of the metric distribution does not change accordingly, as shown by the vague result in \rfig{AlternativeDesign}(d). To solve this problem, we develop the adaptive contour map. In the adaptive map, the diffusion radius is reduced by the same ratio with the map scale when users zoom-in to a finer map granularity. In \rfig{AlternativeDesign}(e), more details on the mobility metric distribution can be visually extracted.

Another challenge on mapping mobility metrics comes from the skewed value distribution. Take the vibrancy metric as an example, a majority of grid can have a similar average vibrancy, i.e., 45\%$\sim$75\% of the maximal vibrancy (\rfig{UrbanFACETOverview}(d)). Then the metric visualization will be similar in color, making it hard to detect distributional patterns. To address this issue, we have applied a double color map filter, as shown in the metric slider of \rfig{UrbanFACETOverview}(g). Two metric value thresholds are set by the filter. The grids below the minimal threshold will be removed in the metric layer, the grids above the maximal threshold will be assigned the rightmost color (i.e., red), the grids in between will be mapped linearly from the newly selected value range to the color palette. Compared with the original visualization result (\rfig{AlternativeDesign}(b)), the new visualization with color map filter (\rfig{AlternativeDesign}(c)) can enhance the stereoscopic depth in color and help to detect more interesting regions. On the color palette, we also support to reverse the color mapping (\rfig{UrbanFACETOverview}(g)), so that the valleys of metric value distribution can be better detected.

To enable fine-grained analysis, we implement three detail views to show the statistics of regions and the associated users. The top view (\rfig{UrbanFACETOverview}(c)) lists the average value of all metrics in the currently selected region. If no region is selected, the entire city's metrics are listed. The middle view (\rfig{UrbanFACETOverview}(d)) depicts the probability density function of grid-based average metric values. This distribution helps users to set the color map filter in the analysis. The bottom view (\rfig{UrbanFACETOverview}(e)) depicts the distribution of users in different POI/DIV classes (for user entropy), and the distribution of users in the current POI/DIV class (for record entropy). These distributions are the raw data to calculate the user mobility metric, and therefore can visually interpret why the mobility metric value becomes high or low in certain regions.

\bsubsec{Interacting for Correlation Analysis}{MobilityVis}

Beyond the static overview, UrbanFACET extends the visual analysis of user-level mobility by correlating these metrics with the side information of location records, such as POIs, regions, record densities, DIV-based demographics and the record time. Four interaction methods are introduced in the visual correlation analysis: overlaid, side-by-side, explicit coding and filtering.

\begin{figure}[t]
\centering
\subfigure[Correlation with house price]{\includegraphics[height=1.8 in]{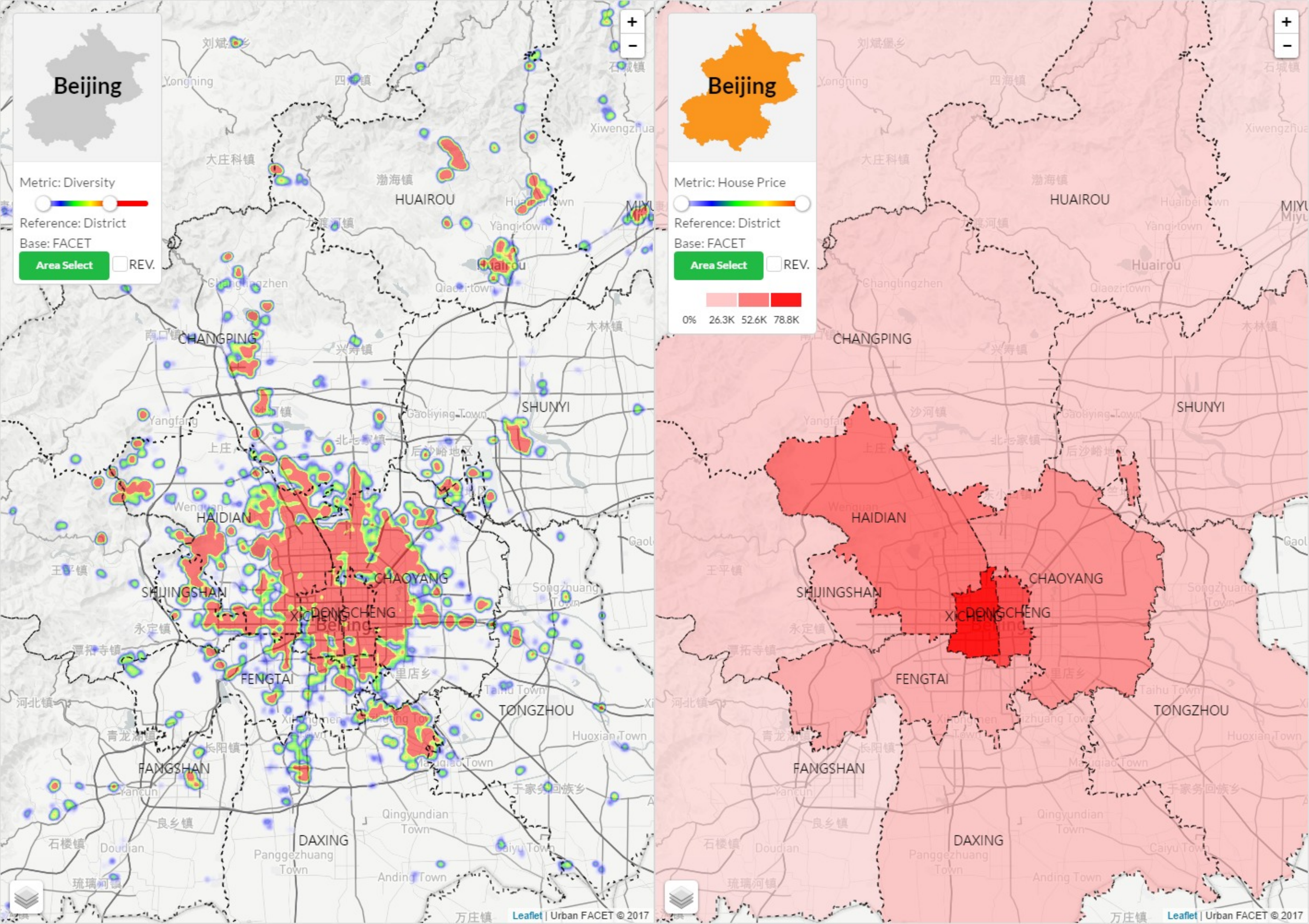}}\\
\vspace{-0.1 in}
\subfigure[Comparison across time of day]{\includegraphics[height=1.8 in]{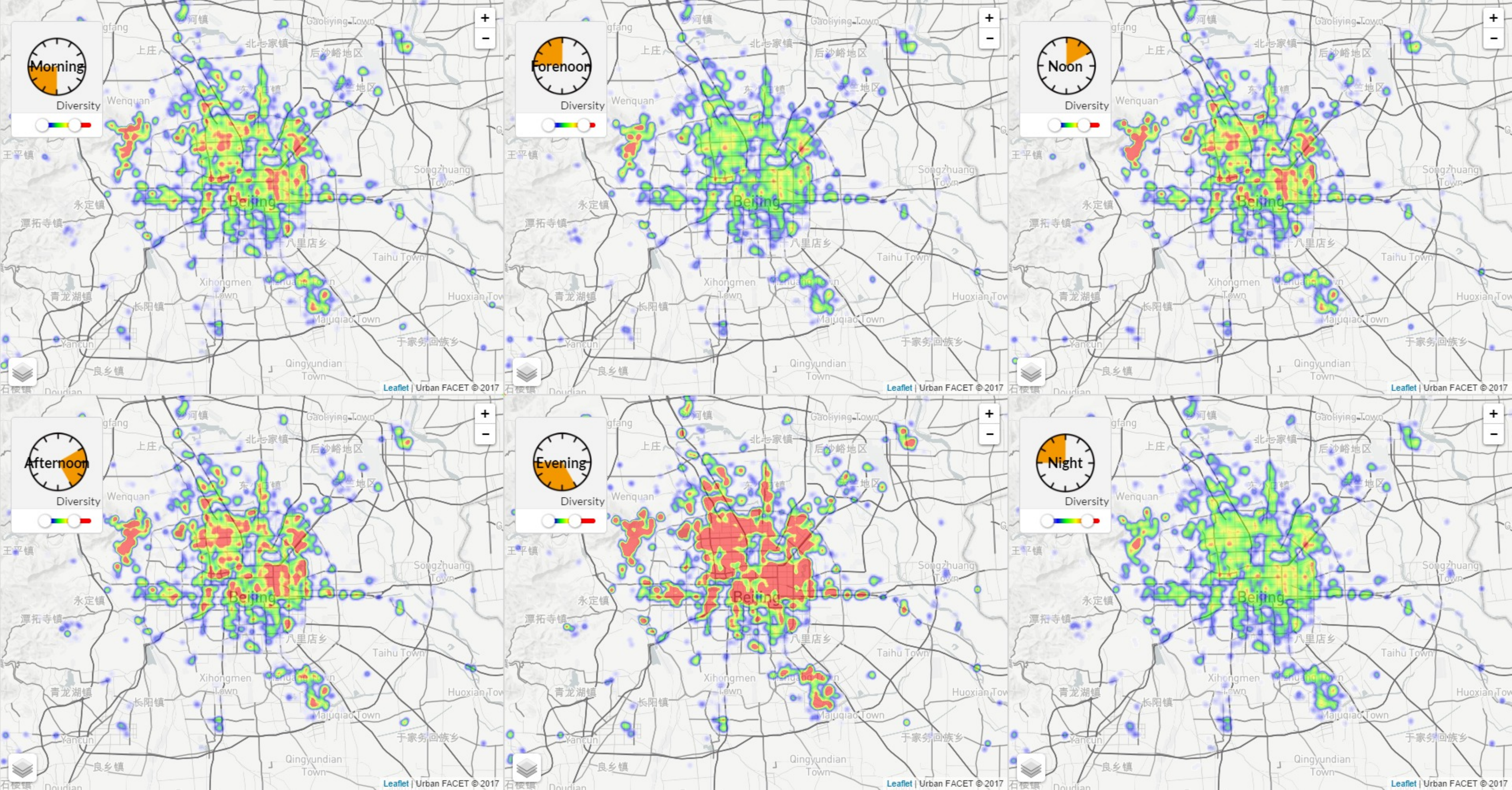}}
\vspace{-0.15 in}
\caption{The diversity metric distribution in Beijing.}
\vspace{-0.2 in}
\label{fig:BeijingDiversity}
\end{figure}

\begin{figure}[t]
\centering
\subfigure[Distribution]{\includegraphics[height=1.2 in]{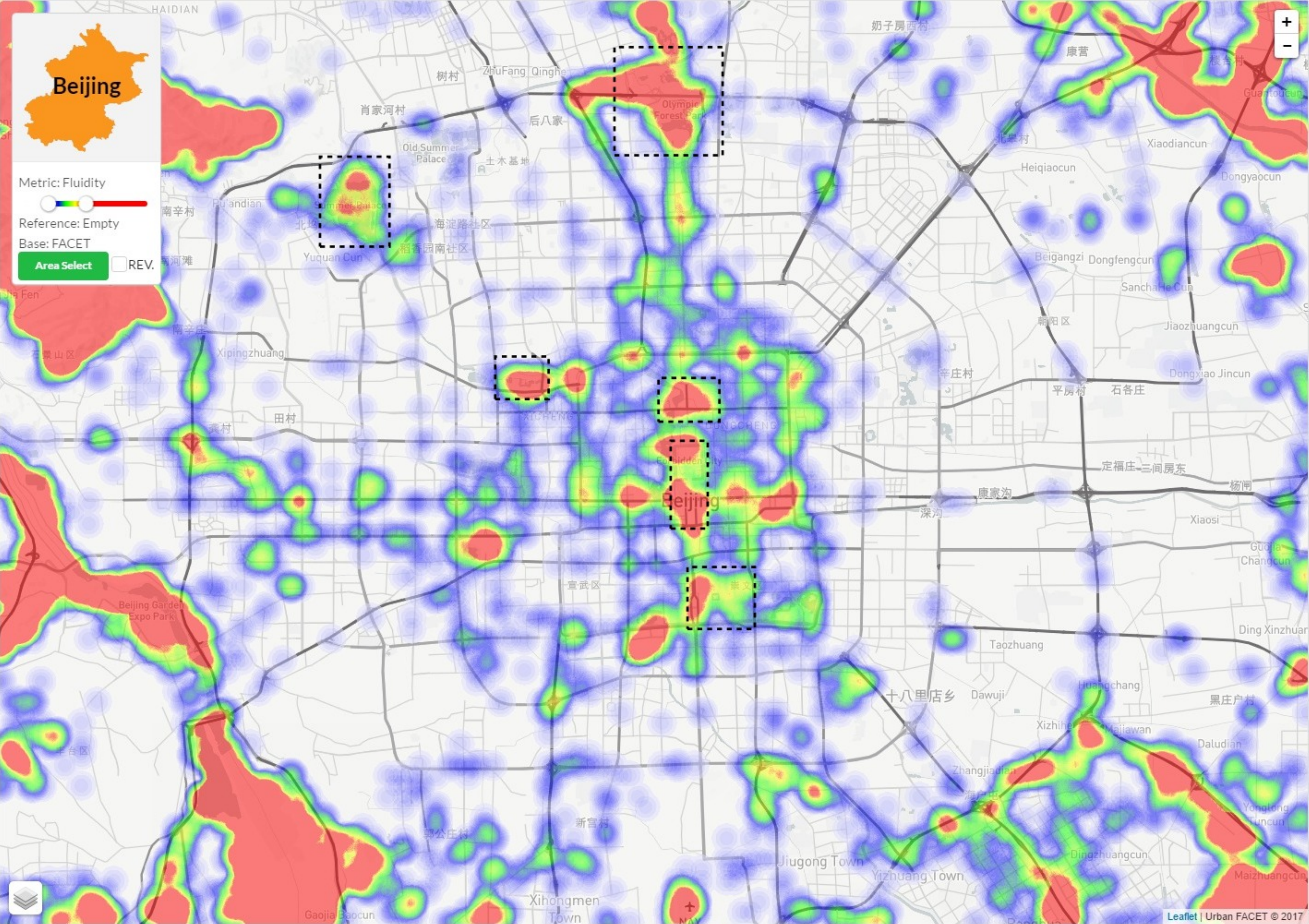}}
\subfigure[Correlation with tourism POIs]{\includegraphics[height=1.2 in]{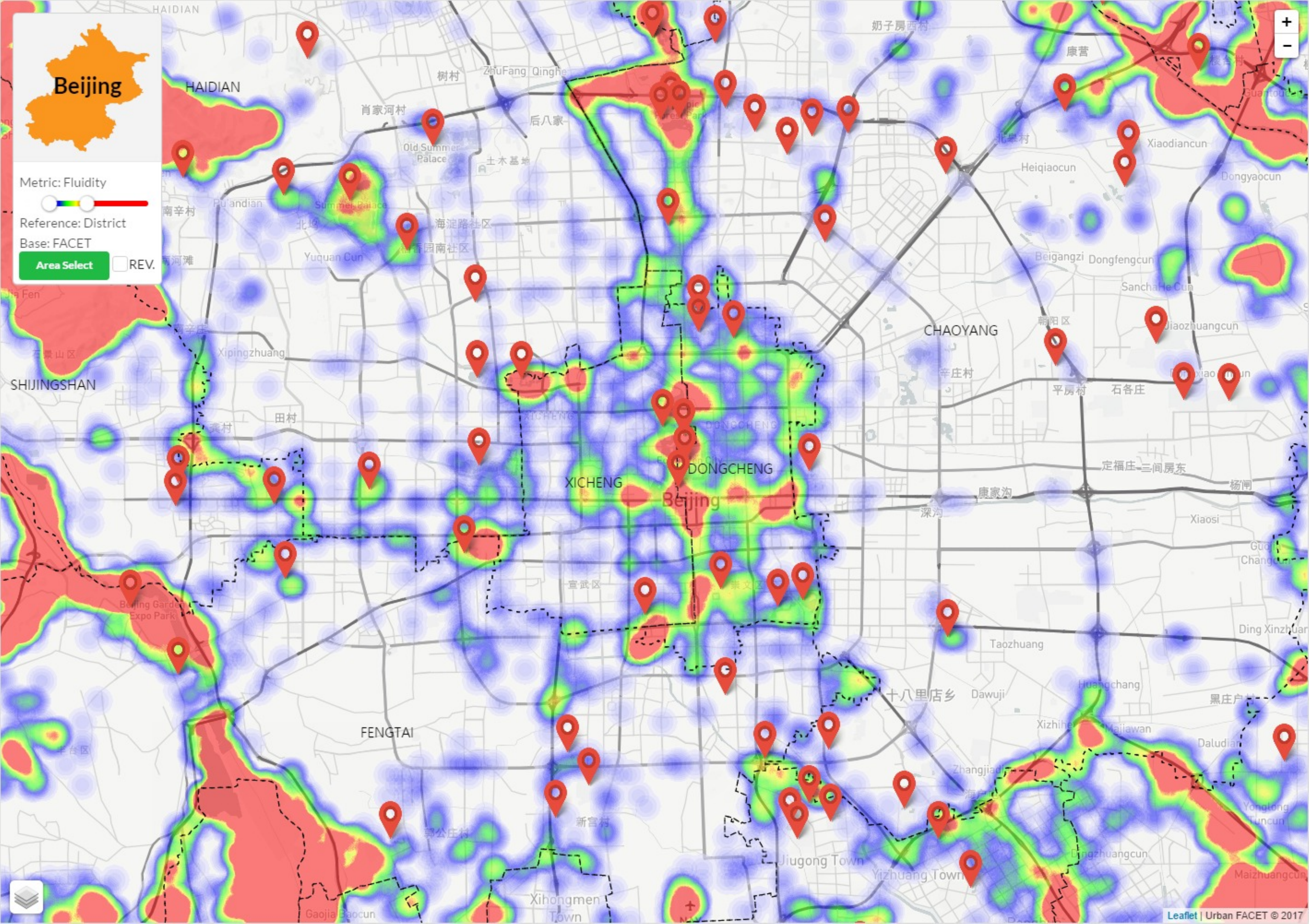}}
\vspace{-0.15 in}
\caption{The fluidity metric distribution in Beijing.}
\vspace{-0.25 in}
\label{fig:BeijingFluidity}
\end{figure}

First, a new reference layer can be overlaid on top of the metric layer by the option in \rfig{UrbanFACETOverview}(b). Two classes of references are introduced, POIs and regions. For POIs, users can select one of the ten classes in \rfig{DataTable}(a) to display. Because the total number of POIs in a city can be overwhelming, we implement a POI filter on the right of the correlation option. Initially, only the POIs within the grids having the highest mobility metric are displayed (\rfig{BeijingFluidity}(b)). Users can set the filter to reveal more POIs correlating with the mobility metric value. For region references, in the coarsest setting, the DIV partition is overlaid on top of the metric layer to show the correlation of DIVs with the metric value (\rfig{BeijingDiversity}(a)). The region granularity can be adjusted by the region controller (in the same place with POI filter) into sub-districts and streets.

\begin{figure}[t]
\centering
\includegraphics[height=1.8 in]{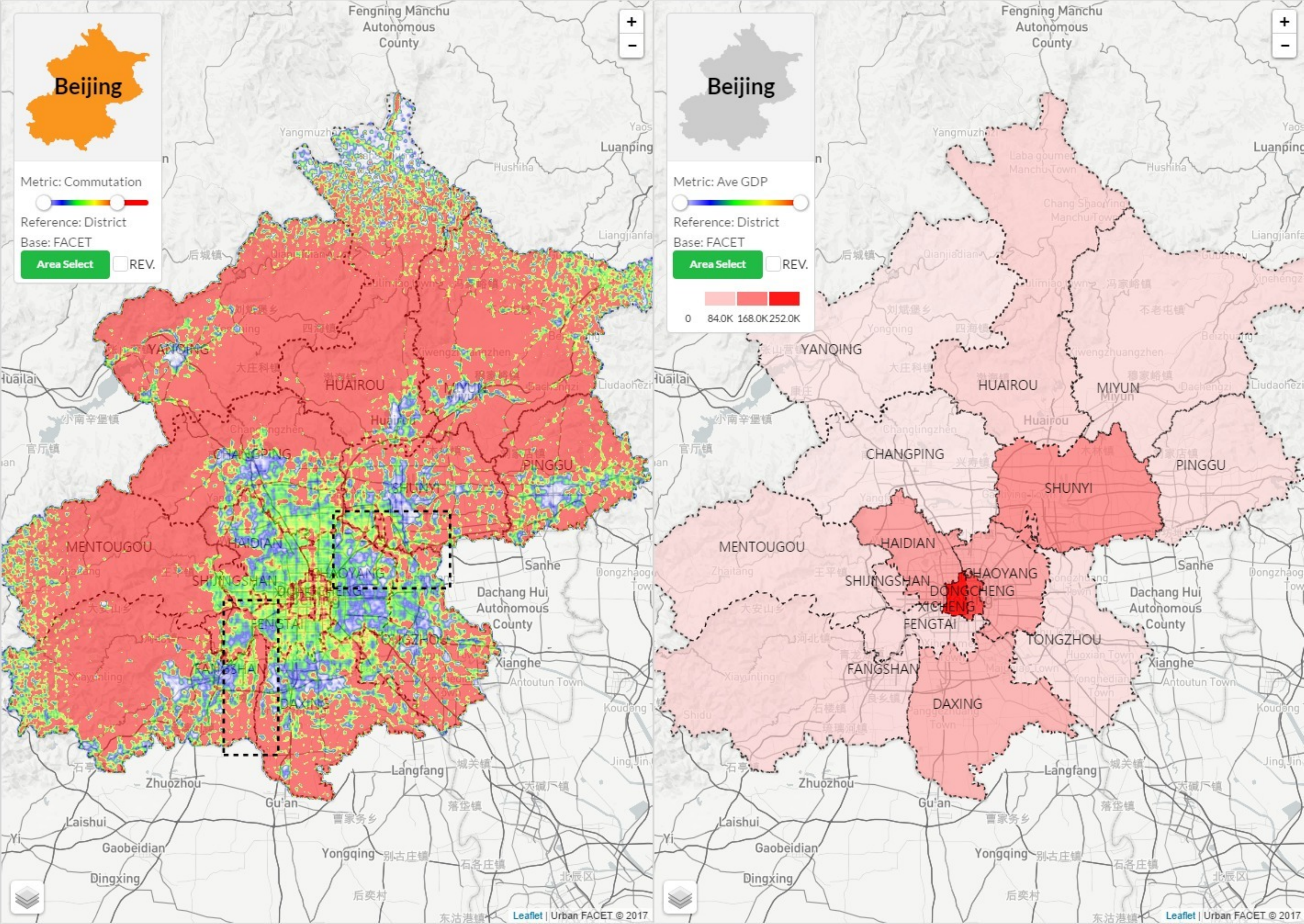}
\vspace{-0.1 in}
\caption{The commutation metric distribution in Beijing.}
\vspace{-0.2 in}
\label{fig:BeijingCommutation}
\end{figure}

Second, the map layer can be split horizontally to add another metric in a side-by-side comparison (\rfig{BeijingCommutation}(b)). This is achieved using the ``+'' button on top of the metric layer configuration in \rfig{UrbanFACETOverview}(b). Both the region demographics (GDP, population, house price) and another mobility metric can be selected for comparison. We also design a suite of interaction methods for the side-by-side comparative analysis, which are detailed in \rsubsec{VisualComp}.

Third, users can choose FACET in the metric layer selector to show ``Fluidity, vibrAncy, Commutation, divErsity, densiTy'' simultaneously. In a star plot design (\rfig{UrbanFACETOverview}(f)), the four mobility metrics are represented at each of the four axes. The record density is mapped to the saturation of the fill color of each star plot (HSL color space). The color hue is fixed to red, with a moderate color lightness. By default, a star plot is drawn on top of each DIV for multidimension comparison. Users can drill-down to sub-district comparison or roll-up to the whole city metrics through the region controller.

Last, time is an important attribute for each record. The mobility metrics can have diversified distribution at different time ranges. Based on the time division in \rfig{DataTable}(b), we support to filter the records according to the time of day or day of week. The mobility metric value is recomputed over the records after time-based filtering (\rfig{BeijingDiversity}(b)).

\bsubsec{Comparing among Time and City}{VisualComp}


UrbanFACET initiates the comparison analysis among time periods (time of day or day of week) and cities by clicking three rightmost buttons in the panel of \rfig{UrbanFACETOverview}(a). On time of day based comparison, all records are divided by six time periods of the same day (ID 0$\sim$5 excluding midnight in \rfig{DataTable}(b)). The spatial metric distribution in each record class is shown in one sub-view of the 3$\times$2 display (\rfig{BeijingDiversity}(b)). Alternatively, by day of week, the records in weekday and weekend can also be compared in a 2$\times$1 display.

We design several interactions for the comparison analysis. First, the zoom and pan operation on each sub-view is synchronized with all the other sub-views in the same display. Users can preserve the spatial context in comparing among sub-views. To focus on small regions, a range selection interaction is introduced. Users can select a rectangle area on one sub-view and get the same area highlighted in all sub-views for examination. Over the contour map, a point selection will retrieve an area with similar metric values. On the same time, the areas with the similar metric value are also highlighted in both the current sub-view and all the other sub-views. This is to facilitate the metric value comparison across time periods.

\begin{figure*}[t]
\centering
\subfigure[Vibrancy]{\includegraphics[height=1.75 in]{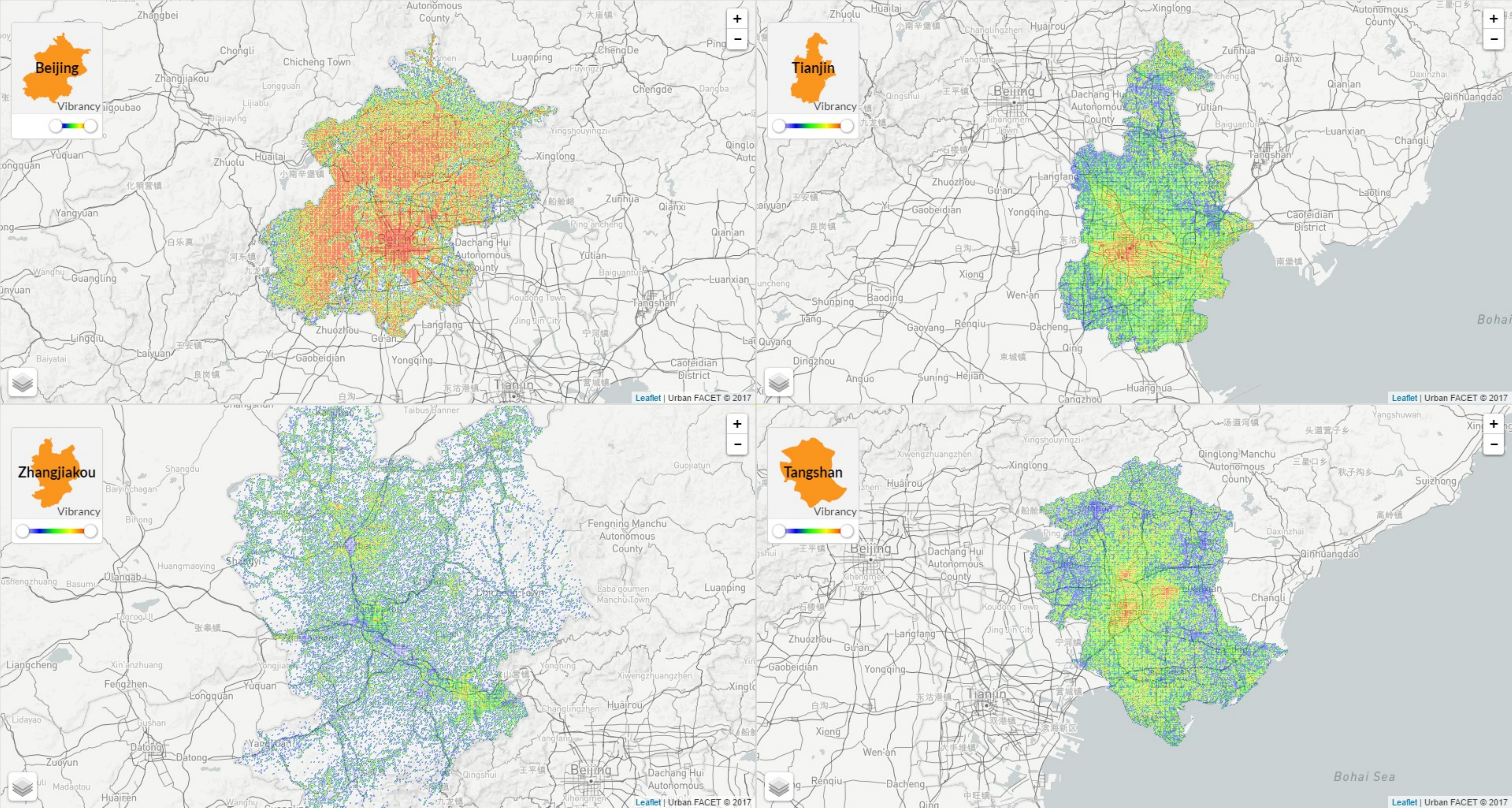}}
\hspace{0.1 in}
\subfigure[Commutation]{\includegraphics[height=1.75 in]{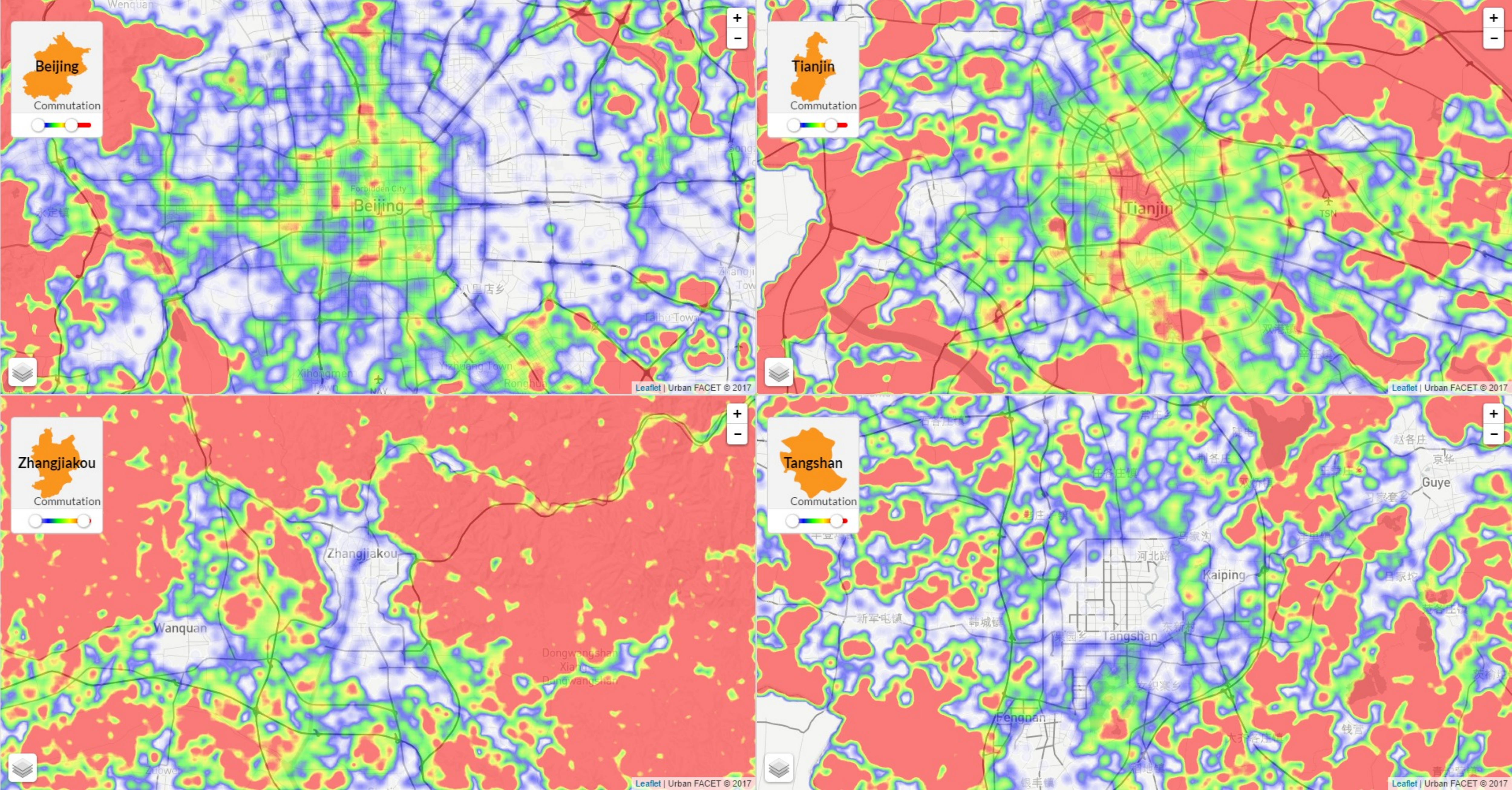}}\\
\vspace{-0.1 in}
\subfigure[Fluidity]{\includegraphics[height=1.75 in]{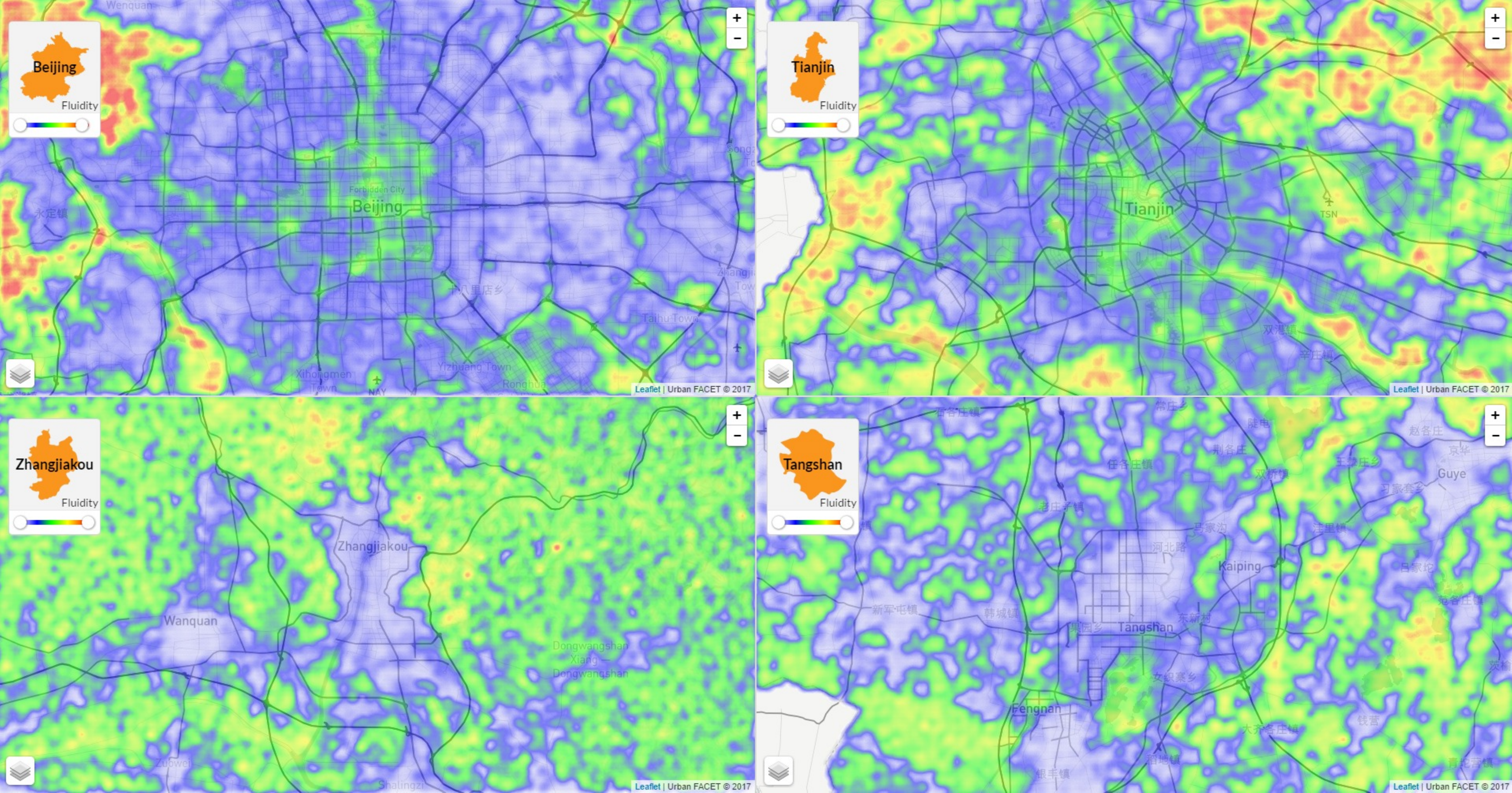}}
\hspace{0.1 in}
\subfigure[Density]{\includegraphics[height=1.75 in]{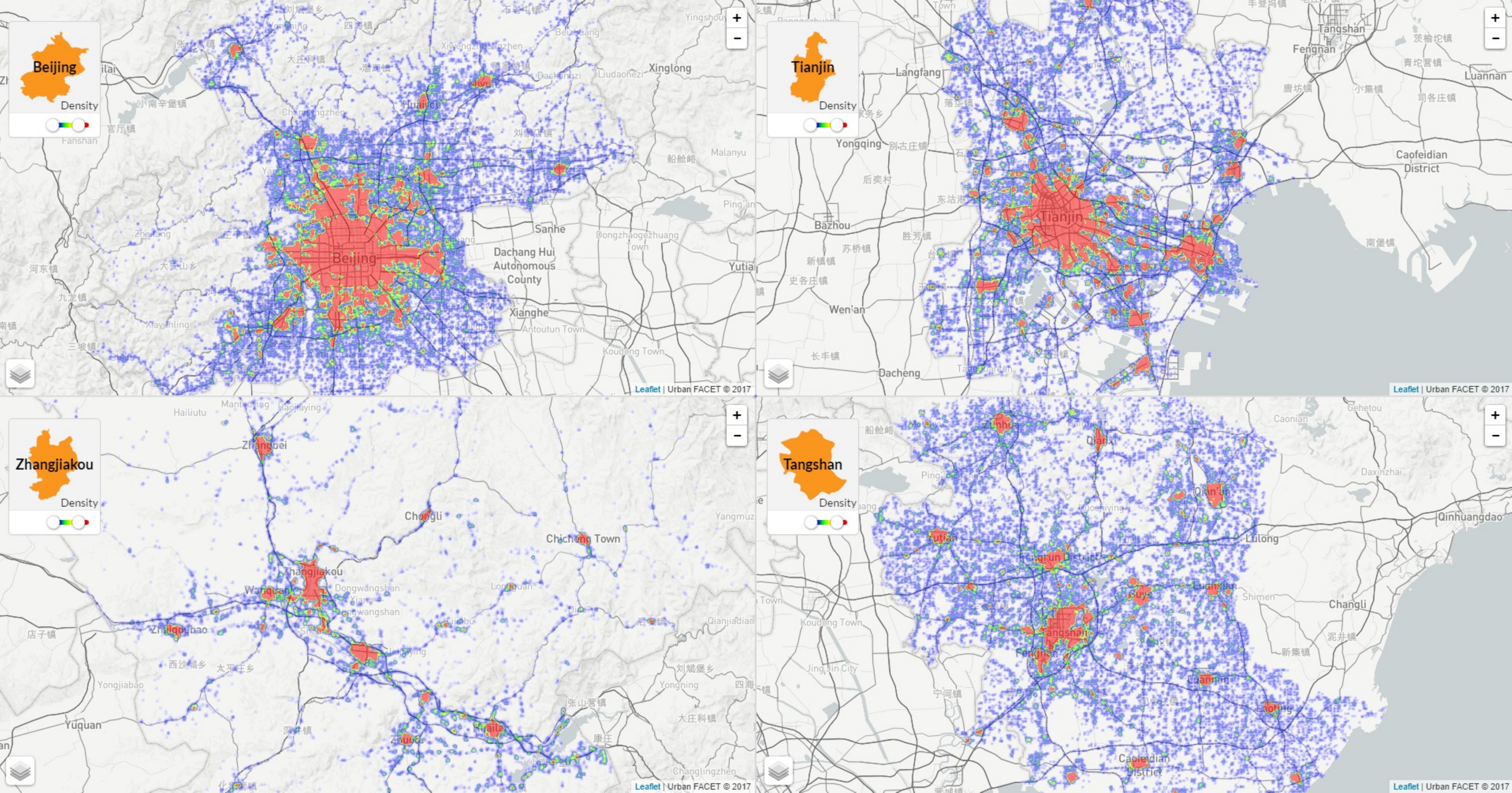}}
\vspace{-0.15 in}
\caption{Mobility metric comparison among Beijing, Tianjin, Tangshan and Zhangjiakou.}
\vspace{-0.2 in}
\label{fig:CityComp}
\end{figure*}

On city comparison, the current version supports up to four cities in a 2$\times$2 sub-view display (i.e., Beijing, Tianjin, Tangshan, Zhangjiakou, as shown in \rfig{CityComp}). Because these cities have different maps, the zoom, pan and range selection operations are not synchronized among sub-views. The point selection interaction to compare value distributions are preserved. 
\bsec{Evaluation}{Eva}

\bsubsec{Case Study}{Case}

We have deployed UrbanFACET to study the four representative cities in China (\rtab{DataCollection}). Before this study, we conducted an interview with the urban data experts from the mobile analytics company, whose everyday job is to apply this data set to solve industry problems, including traffic optimization, facility configuration and crowd analysis for commerce. After pilot usages of UrbanFACET on the selected cities, the experts give detailed comments on the real-life tasks well supported by our system. First, while existing data analytics are mostly targeted to optimize the efficiency of certain industry, little attention has been made to urban users who play a central role in urban industries. UrbanFACET has the potential to foresee the everyday life of these users, characterize the profile of cities where they live, and explore the interplay of cities and their residents. Second, on the detailed tasks, UrbanFACET can help to plan the location of public facilities and optimize precision advertising, by demand analysis via urban user mobility demographics. Third, the unification of national capital region becomes a recent hot topic in urban planning industry of China. Profiling and comparing these cities provides key values to the coordinated development of Beijing and its surrounding areas.


\bsubsubsec{Beijing}{Beijing}

\begin{figure}[t]
\centering
\subfigure[Full scale]{\includegraphics[height=1.3 in]{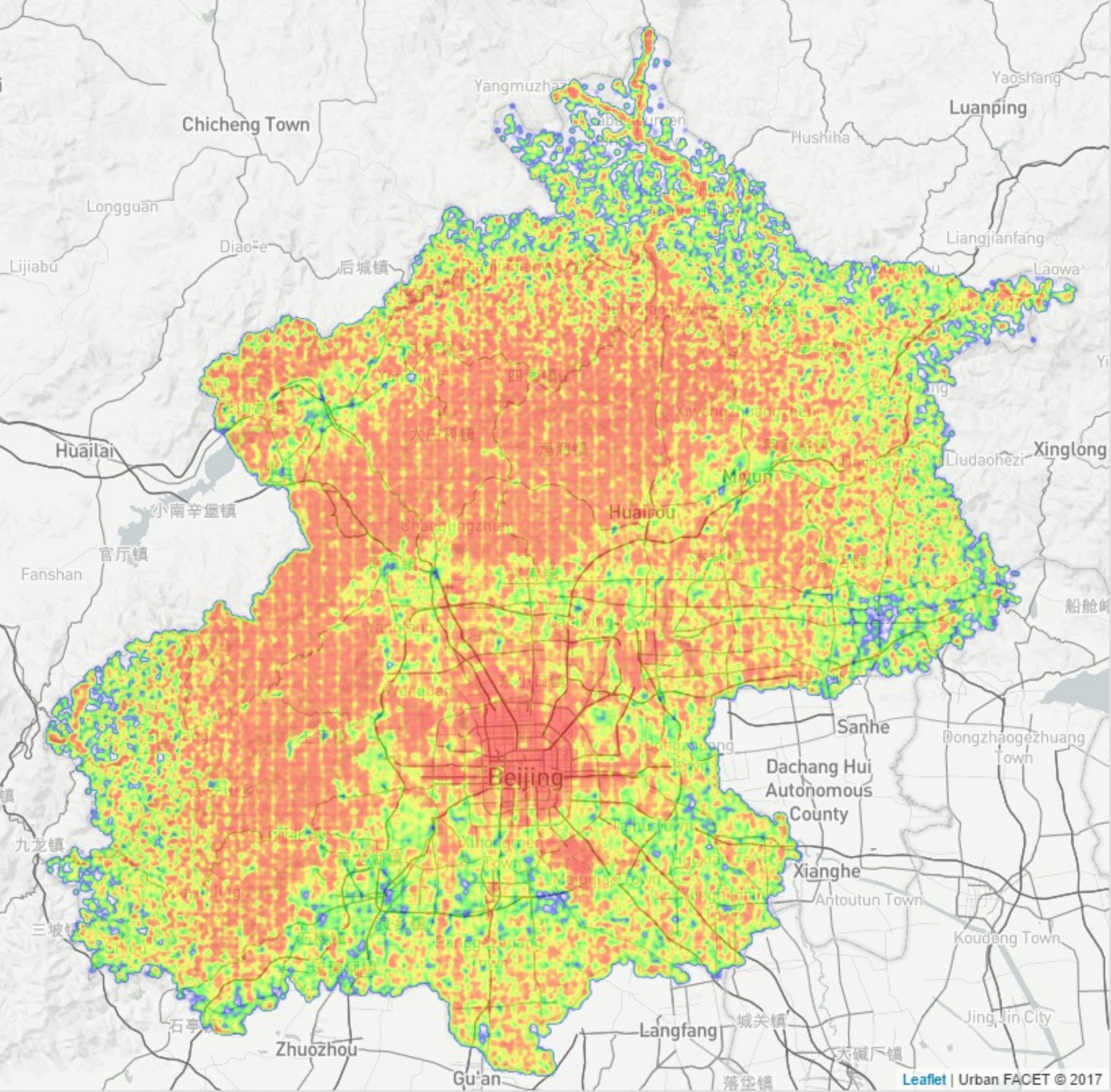}}
\hspace{0.02 in}
\subfigure[Terrain map]{\includegraphics[height=1.3 in]{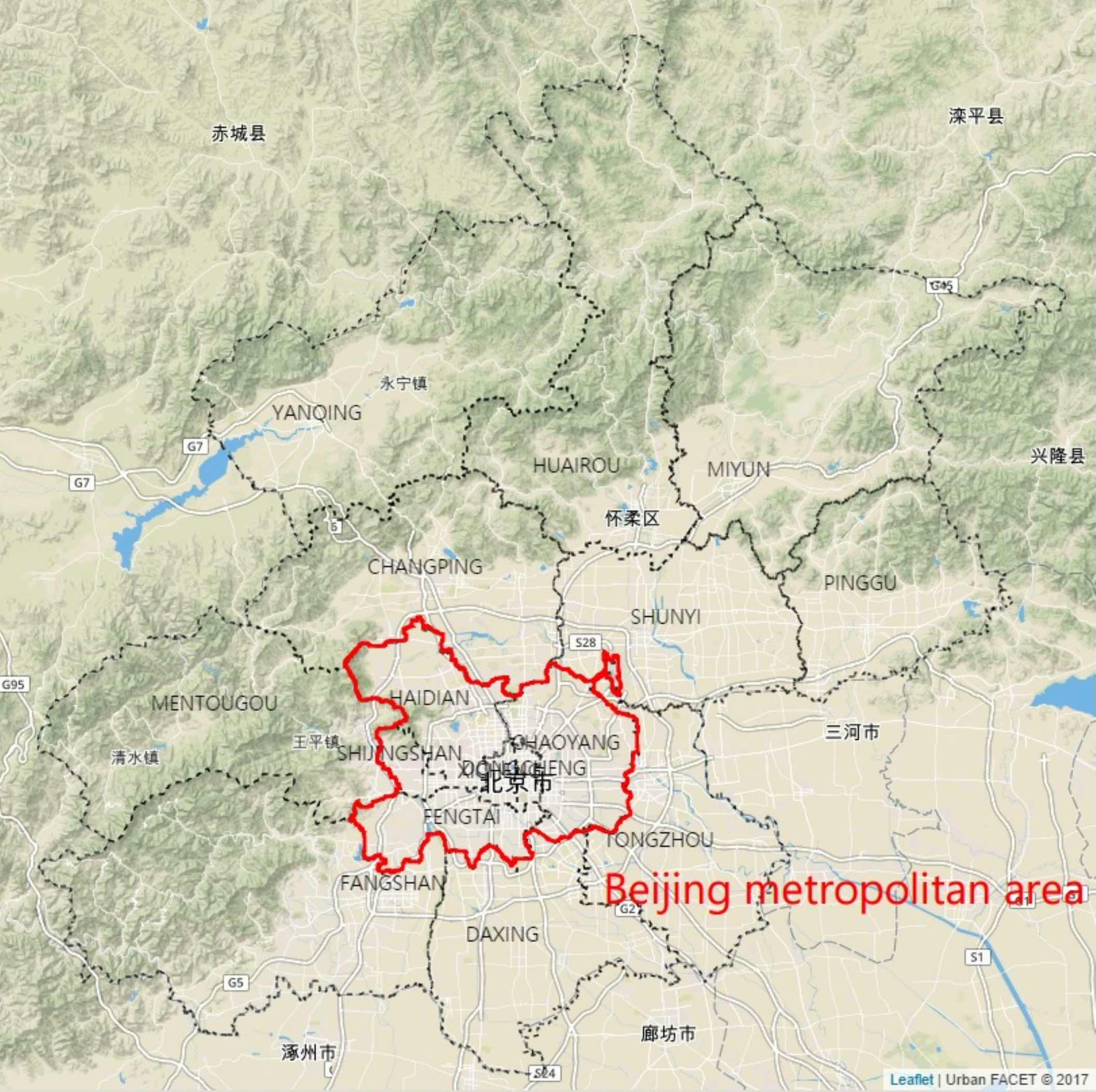}}
\vspace{-0.15 in}
\caption{The vibrancy metric distribution in Beijing.}
\vspace{-0.1 in}
\label{fig:BeijingVibrancy}
\end{figure}

In this part, we conduct a detailed analysis of the four mobility metrics over Beijing.

\textbf{Vibrancy:} As discussed in \rsubsec{UserMetric}, this metric indicates the citizen's overall life abundance. A high vibrancy means s/he gets accessed to many classes of POIs uniformly. The overall vibrancy distribution in Beijing is shown in \rfig{BeijingVibrancy}(a). By comparing with the terrain map in \rfig{BeijingVibrancy}(b), two high vibrancy regions can be found: a) the center of Beijing, which is the metropolitan area (metro in short); b) the western, northern and northeastern mountain regions, where most records are made by travelers lived in Beijing metro area. As we can see in \rfig{CityComp}(d), the mountain regions of Beijing is very low-populated.

We drill down to more details by zooming into the metro area, which is shown in \rfig{UrbanFACETOverview}(f). Traditionally, the central city of Beijing is partitioned by six ring-shaped roads, and is developed radially. The population density is almost symmetric with Tiananmen Square as the center (\rfig{CityComp}(d)). However, from the vibrancy distribution in \rfig{UrbanFACETOverview}(f), we observe a different, asymmetric pattern that the hotspots roughly locate in a rectangle: ring-5.5 in the north, ring-3 in the south, ring-4 in the east and ring-3.5 in the west. In other words, the northern people lives a more abundant life than the southern; eastern people is a little better than western. Outside the metro area, we also notice a few high vibrancy area: the extended line of Chang'an Avenue, Xiangshan park in the northwest, Beijing Economic-Technologlcal Development Area in the southeast, and the Capital International Airport in the northeast. 
All these findings correspond well with the latest status of Beijing by consulting with senior citizens there.

\textbf{Commutation:} The commutation map of Beijing is shown in the left part of \rfig{BeijingCommutation}. There is a clear pattern of three radial layers. People in the regions outside the metro area need to commute to other regions (i.e., the metro area) to meet their demands. People in the metro area but outside the city center has a low commutation value, who mostly can satisfy their everyday requirement within their own region. People in the center of Beijing (i.e., inside the third ring) also need to commute more. We hypothesize, this is because most people do not live in the city center, but only work or visit for routine affairs in the government departments. Our hypothesis can be validated by the line and circular hotspots overlapping with subway line 1, 2 and 8 in the city center. The line hotspot in the southwest of the metro area is found to be the National Route 107, the busiest road in China that connects northern and southern China. The line hotspot in the northeast is found to be Beijing-Hangzhou Grand Canal.

We compare the commutation map with other side information. In the right part of \rfig{BeijingCommutation}, GDP is found to be negatively correlated with the commutation metric. The richer an area, the fewer commutation is required for their people.

\textbf{Diversity:} The diversity distribution in Beijing is similar to vibrancy, as shown in the left part of \rfig{BeijingDiversity}(a), except that the mountain regions are not diversified. It is because this metric is based on the region's property, despite of the vibrancy of visitors, the mountain region does not provide many POIs to visit. In the right part of \rfig{BeijingDiversity}(a), the diversity metric is shown to be positively correlated with the house price distribution. More diversified areas get accessed to more abundant life style, thus raises the price of local real estate. In the time-based comparison of \rfig{BeijingDiversity}(b), it can be noticed that the city is less diversified in the early morning and the late midnight, when most people are probably at home. The highest diversity happens in the evening, when a large number of citizens will have some entertainments after their daily work.

\textbf{Fluidity:} In the fluidity distribution, we detect the pattern of three radial layers similar to those in the commutation metric (\rfig{BeijingCommutation}). Compared to commutation, fluidity focuses more on the region itself. After examining every hotspot in the metro area of \rfig{BeijingFluidity}(a), we find that these high-fluidity regions correlate perfectly with famous tourist attractions in Beijing: Tiananmen, The Forbidden City, The Temple of Heaven, Shichahai, The Olympic Park, The Beijing Zoo, The Summer Palace, etc. The correlation becomes more salient when we overlay tourism POIs as the reference layer (\rfig{BeijingFluidity}(b)).

We also overlay the star plot synthesizing all four mobility metrics and the population density over each DIV in \rfig{UrbanFACETOverview}(f). The DIVs in the city center (DONGCHENG, XICHENG), metro areas outside the center (CHAOYANG, HAIDIAN, etc.), and remote DIVs (SHUNYI, TONGZHOU, DAXING, etc.) are quite different with each other.




\bsubsubsec{Tianjin}{Tianjin}



%

\begin{figure}[t]
\centering
\subfigure[Vibrancy]{\includegraphics[height=1.2 in]{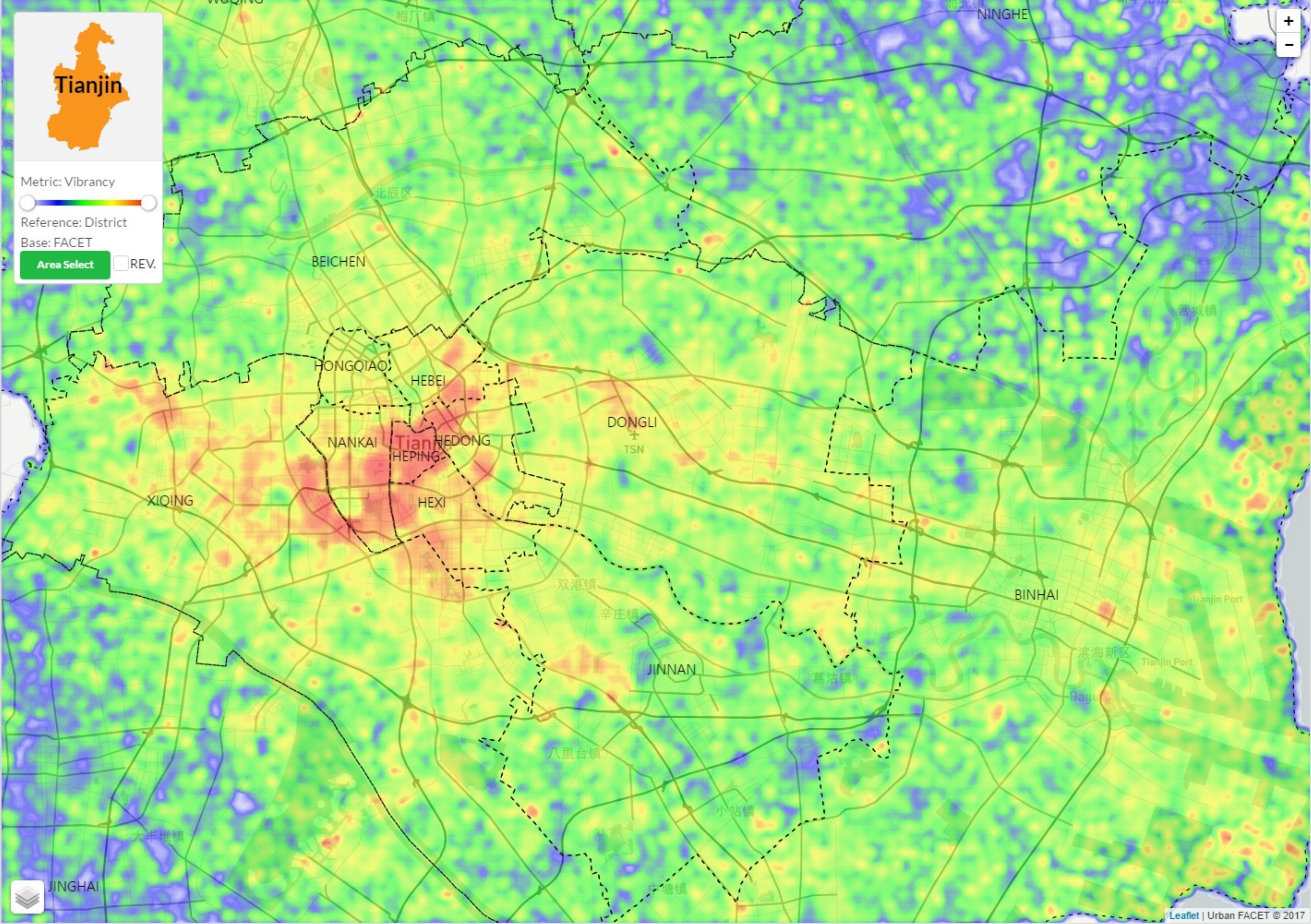}}
\vspace{0.02 in}
\subfigure[Commutation]{\includegraphics[height=1.2 in]{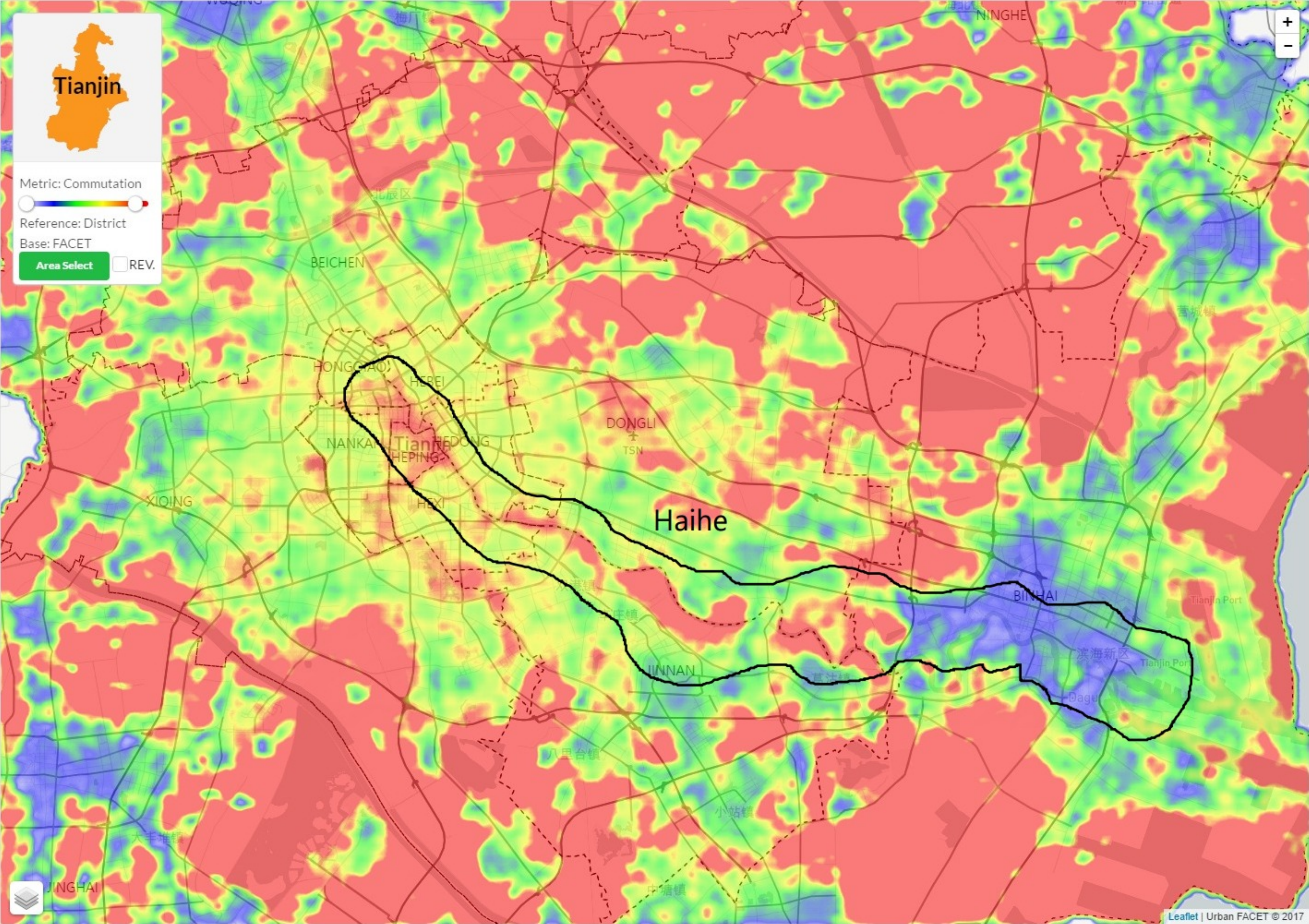}}
\vspace{-0.2 in}
\caption{The mobility metric distribution in Tianjin.}
\vspace{-0.2 in}
\label{fig:TianjinMobility}
\end{figure}

Tianjin, as the top five cities in China, is similar to Beijing in many ways. Below, we will report our analysis using UrbanFACET. On the other hand, their city-level differences are described in \rsubsec{CityComp}.


\textbf{Vibrancy:} In \rfig{TianjinMobility}(a), the vibrancy distribution in Tianjin is shown to be centric to the Heping DIV, which is the economic and financial center of the city. The four DIVs of Nankai, Hebei, Hedong and Hexi surrounding Heping DIV also maintain a moderate level of vibrancy, forming a first ring. In the suburb, there are four satellite DIVs in the second ring which are less vibrant (Xiqing, Beichen, Dongli and Jinnan). In comparison, Xiqing DIV has a larger vibrancy because of its role as the new economic development zone of the city.

\textbf{Commutation:} The commutation distribution of Tianjin (\rfig{TianjinMobility}(b)) is also central to Heping DIV. The major difference lies in that the high commutation zone stretches from northwest to southeast and southwest, while the high vibrancy zone is from northeast to southwest. This distribution fits well with the population and geography pattern of Tianjin. First, Nankai DIV hosts most famous universities in Tianjin. The students there have less commutation than other citizens, therefore creates the low commutation valley in the left metro area of Tianjin. Second, the major river across Tianjin, i.e., the Haihe river as annotated in \rfig{TianjinMobility}(b), runs from northwest to southeast. Citizens will mostly move along the river as crossing it can take a larger time cost.

\bsubsubsec{Comparison among Cities}{CityComp}


We also compare the four cities on four mobility metrics. \rfig{CityComp}(a) shows the vibrancy metric: Beijing, as the capital city of China, enjoys an almost full coverage of abundant life for people within the administrative boundary; in comparison, Tianjin (city center and coastal new city) and Tangshan (city center) only have small areas having vibrant people; on the other hand, the city of Zhangjiakou is relatively weak in life abundance.

On commutation, as shown in \rfig{CityComp}(b), the people in Beijing only need to commute more if s/he lives in the city center (within the third ring), the outsider people can meet their demands in the local DIV. In Tianjin, the city planning seems to be not as good as Beijing, more people live/work in the city center and needs to commute more across DIVs. This higher transportation cost than Beijing is acknowledged by the local citizen in Tianjin. Tangshan and Zhangjiakou are similar in that they mostly live in the city center and they do not need to commute, i.e., probably close to their working site.


On fluidity, which indicates the tourist attractions. \rfig{CityComp}(c) shows that Beijing has many hotspots for tourism in the city center, so does Tianjin, while Tangshan and Zhangjiakou barely have tourist attractions in the city region. Lastly, we also compare their record density distributions. As shown in \rfig{CityComp}(d), the density mostly radiates out from the center of each city. This distribution is rather different from the four mobility metrics, which demonstrates the value of our user-level mobility analysis work.

\bsubsec{User Experiment}{User}



We conducted a controlled experiment to understand the novelty of user-level mobility metrics in comparison to the classical record density metric. For each of the five metrics, UrbanFACET with the same color-metric mapping function is used.

\textbf{Experiment design.} We recruited 16 subjects (12 male, 4 female), all were senior students or engineers living in Beijing for more than three years. The experiment was divided into the training session and the test session for each visualization. In the training session, users warmed up by completing the same suite of tasks on a different city (e.g., Tangshan). The organizer checked the result of each training task and addressed all questions before proceeding to the next step.

\textbf{Task.} Each subject completes eight tasks, which are divided into four task groups. Most tasks are for Beijing unless otherwise stated.

\emph{TG1 (Vibrancy): a) Judge the degree of metric value symmetry in Beijing metro area; b) Estimate the coverage rate of three hotspots (Zhongguancun, Wangjing, CBD) in all high value area.}

\emph{TG2 (Fluidity): a) Infer the type of high metric value, from residential areas (i.e., low fluidity) to attractions (i.e., high fluidity); b) Estimate the coverage rate of high metric value with inter-city routes.}

\emph{TG3 (Commutation): a) Judge the degree of negative correlation between the metric value and local GDP; b) Judge the degree of correlation between the metric value and DIV centers.}

\emph{TG4 (Diversity): a) Estimate the number of high-diversity sites (e.g., airport) covered by the high metric value area; b) Judge the nonuniform degree of the metric distribution in metro area (Tianjin).}

Note that the two tasks in the same group share the same purpose. The experiment applied a between-subject design. For each subject, s/he will be arranged to complete one task in a group with the mobility metric visualization, and the other task in the same group with the density visualization. The pairing of visualization with task is shuffled for different subjects to eliminate learning and ordering effect. For each task, multiple numbered choices are designed for selection. The ground truth is, the mobility metric distribution will lead to a larger number, while the density distribution will has a smaller number. We recorded the subject's answer after the test. 

\begin{figure}
\centering
\includegraphics[height=1.5 in]{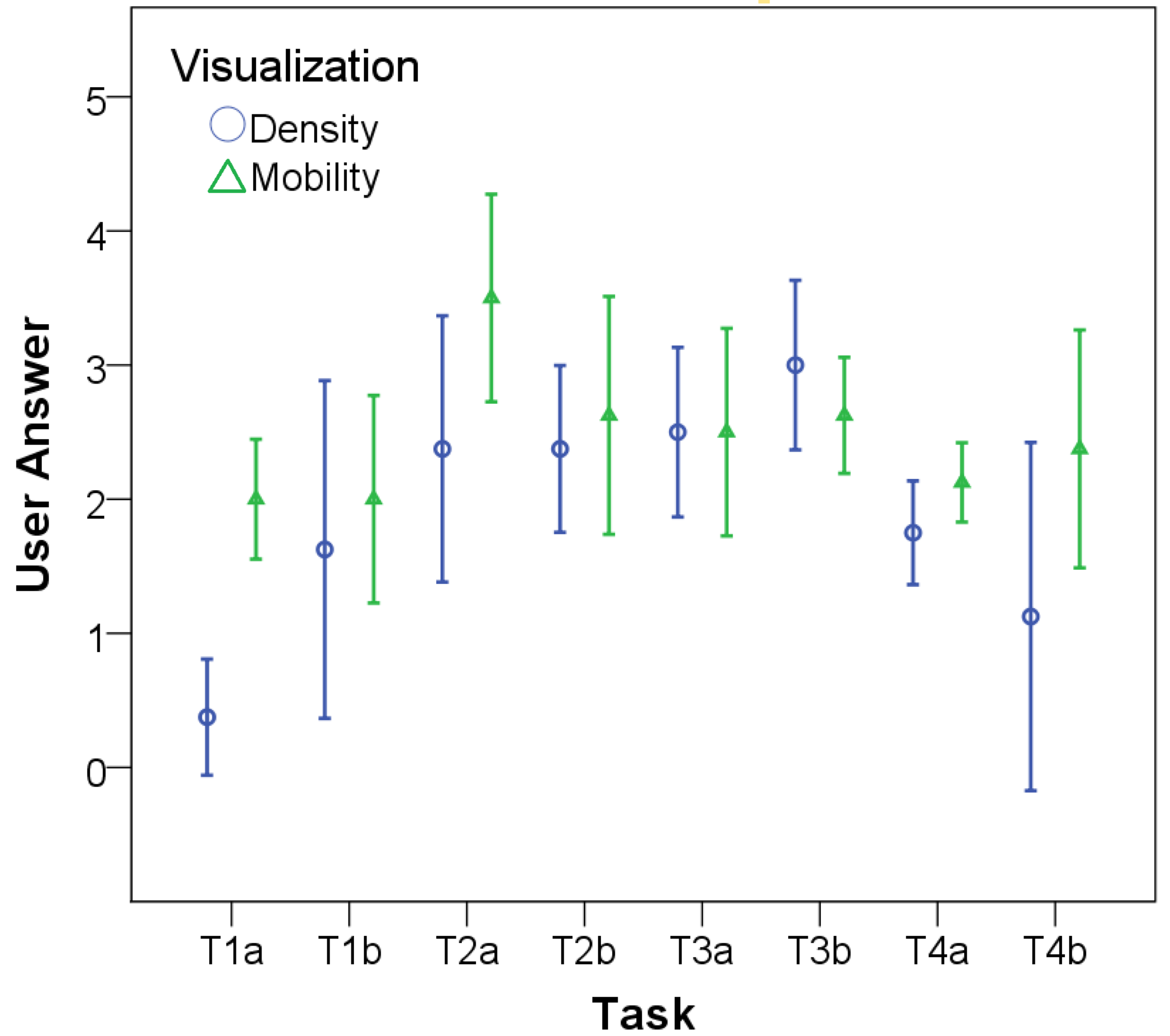}
\vspace{-0.15 in}
\caption{Distribution of user answers in eight tasks.} \label{fig:TaskAnswer}
\vspace{-0.18 in}
\end{figure}

\textbf{Result.} Experiment results were analyzed separately for each task. Significant level was set at 0.05 throughout the analysis. The subject's answers are summarized in \rfig{TaskAnswer}. It is observed that, the mobility metric visualization leads to a larger choice in seven tasks except T3b\footnote{A lot of subjects complain that they are not sure where are DIV centers, which may explain this result.}, which indicates more novel information obtained. We then apply the Mann-Whitney test to analyze the difference between two visualizations, which does not require a normality assumption of the observed answer. It is shown that, between the mobility and density metric, the answer is significantly different on T1a ($U = 1.5, p < .001$), close to significant on T2a ($U = 15.0, p = .083$) and T4b ($U = 15.0, p = .083$), not significant on other tasks. The user study results demonstrate, the four user-level mobility metrics in most cases introduce new information for users (i.e., Vibrancy-T1, Fluidity-T2, Diversity-T4).


\section{Conclusion}

We present a metropolitan-scale visual analytics study over a new urban movement data collection in China. Compared with previous works, our data set is immensely huge, records user's comprehensive movement pattern, and preserve a clear user context for study. To efficiently process, visualize and analyze such big urban movement data, we have proposed: 1) a scalable, grid-based data analytics pipeline to extract and represent user mobility data; 2) a suite of information-theory based metrics to effectively characterize user-level mobility patterns; 3) an integrated visual analytics system, namely UrbanFACET, to support interactive analysis on the multifaceted mobility metrics and their correlation with urban structure and POI distributions, as well as to compare across urban regions and among heavily populated cities.

In future, we plan to extend UrbanFACET in three directions. First, besides the macro-level city profiling, it is also important to model fine-grained regions and conduct analysis on local residents. Second, beyond the spatial feature based entropy metrics, we are interested to define time-based user mobility metrics given the non-uniform temporal distribution of user's movement records. Third, visual storytelling with multifaceted urban movement data poses a demanding challenge to extend the user base of UrbanFACET beyond domain experts. 



%

\bibliographystyle{abbrv-doi}

\bibliography{UrbanFACET}
\end{document}